\documentclass[journal]{IEEEtran}

\usepackage[pdftex]{graphicx}
\graphicspath{{Figs/}}

\usepackage{comment}
\usepackage{caption}
\usepackage{xcolor}
\usepackage[shortlabels]{enumitem}
\usepackage{subcaption}
\usepackage{multirow}

\usepackage{mathptmx}       % selects Times Roman as basic font
\DeclareMathAlphabet{\mathcal}{OMS}{cmsy}{m}{n}
\usepackage{helvet}         % selects Helvetica as sans-serif font
\usepackage{courier}        % selects Courier as typewriter font
\usepackage{type1cm}        % activate if the above 3 fonts are
\usepackage{makeidx}         % allows index generation

\usepackage{amsmath,amsfonts,amssymb,amscd,amsthm,xspace}
\usepackage{mathtools}

\usepackage{epsfig}
\usepackage{enumitem}
\usepackage{bbm}
\usepackage{dblfloatfix}
\usepackage{bm}
\usepackage{booktabs}
\usepackage{color}

\usepackage[ruled,algo2e]{algorithm2e}
\usepackage{algorithm}
\usepackage{algpseudocode}

\makeatletter
\newenvironment{myprocedure}[1][htb]{%
    \renewcommand{\ALG@name}{Procedure}% Update algorithm name
   \begin{algorithm}[#1]%
  }{\end{algorithm}}
\makeatother

\newtheorem{theorem}{\textit{Theorem}}
\numberwithin{theorem}{section}

\numberwithin{corollary}{section}

\newtheorem{assumption}{\textit{Assumption}}

\newtheorem{definition}{\textit{Definition}}
\theoremstyle{definition}
\numberwithin{definition}{section}
\numberwithin{example}{section}

\newcommand{\nt}{T}
\newcommand{\nm}{m}
\newcommand{\nn}{n}

\DeclareMathOperator*{\argmin}{arg\,min}

\newcommand{\norm}[1]{\left\lVert#1\right\rVert}
\newcommand{\MPC}{\mathsf{MPC}}
\newcommand{\NN}{\mathsf{NN}}

\newcommand{\ouralg}{{\textsc{OOD-Charging}}\xspace}

\newcommand{\decayfactor}{\lambda}

\newcommand{\bigone}{\mbox{$I$}}
\newcommand{\bigzero}{\mbox{$0$}}
\newcommand{\rvline}{\hspace*{-\arraycolsep}\vline\hspace*{-\arraycolsep}}

\begin{document}

\title{Out-of-Distribution-Aware Electric Vehicle Charging}

\author{Tongxin~Li,
~\IEEEmembership{Member,~IEEE}, Chenxi~Sun~\IEEEmembership{Member,~IEEE}
\thanks{
This work was supported by the National Natural Science Foundation of China under Grant No. 72301234, 62336005, the Pengcheng Peacock Supporting Scientific Research Fund Category C (2024TC0024), the Guangdong Provincial Key Laboratory of Mathematical Foundations for Artificial Intelligence (2023B1212010001), the Shenzhen Key Lab of Crowd Intelligence Empowered Low-Carbon Energy Network (No. ZDSYS20220606100601002), and the startup funding UDF01002773 of CUHK-Shenzhen.
(\textit{Corresponding
Author: Tongxin Li}).}
\thanks{Tongxin Li is with the School of Data Science (SDS), The Chinese University of Hong Kong, Shenzhen, China (e-mails: litongxin@cuhk.edu.cn)}
\thanks{Chenxi Sun is with the Shenzhen Institute of Artificial Intelligence and Robotics for Society, Shenzhen, China (email: sunchenxi@cuhk.edu.cn)}
}

\maketitle
\IEEEpeerreviewmaketitle

\begin{abstract}
We tackle the challenge of learning to charge Electric Vehicles (EVs) with Out-of-Distribution (OOD) data. {Traditional scheduling algorithms typically fail to balance near-optimal average performance with worst-case guarantees, particularly with OOD data. Model Predictive Control (MPC) is often too conservative and data-independent, whereas Reinforcement Learning (RL) tends to be overly aggressive and fully trusts the data, hindering their ability to consistently achieve the best-of-both-worlds.} To bridge this gap, we introduce a novel OOD-aware {scheduling algorithm, denoted \ouralg.} This algorithm employs a dynamic ``awareness radius'', which updates in real-time based on the Temporal Difference (TD)-error that reflects the severity of OOD. The \ouralg algorithm allows for a more effective balance between consistency and robustness in EV charging schedules, thereby significantly enhancing adaptability and efficiency in real-world charging environments. Our results demonstrate that this approach improves the scheduling reward reliably under real OOD scenarios with remarkable shifts of EV charging behaviors caused by COVID-19 in the Caltech ACN-Data.

% the trustworthiness of the machine-learned policy
% we provide a learning-augmented policy that uses a robustness budget, learned online from the temporal difference error of a reinforcement learning policy, to infer the trustworthiness of the machine-learned policy and therefore balance a trade-off between consistency and robustness.

\end{abstract}



% \fi

\IEEEpeerreviewmaketitle

\section{Introduction}
\label{sec:intro}

In the era of the Electric Vehicle (EV) boom, developing efficient charging solutions to optimize EV operations has emerged as a critical challenge, especially for renewable-integrated EV charging stations~\cite{chen2017dynamic,yan2018optimized}. This adaptive charging problem becomes more complex due to uncertainties induced by both renewable generation and human behavior. For instance, the widespread installation of rooftop photovoltaic (PV) solar systems at workplace charging stations enhances sustainability. However, it also introduces variability in the dynamics of a solar-powered charging system. Non-commercial EVs, with their heterogeneous charging profiles determined by human drivers and battery specifications, further compound the uncertainty in charging dynamics.

To deal with the induced uncertainties, recent times have seen a growing interest in utilizing data-driven and machine learning methods to develop advanced charging schemes~\cite{chics2016reinforcement,wan2018model,li2019constrained,li2021learning,fachrizal2022optimal,li2022ev}. In experimental simulations, these techniques have demonstrated superior performance over traditional scheduling algorithms, such as Earliest Deadline First (EDF), Least Laxity First (LLF), and predictive scheduling~\cite{lee2021adaptive}, particularly when these conventional methods struggle to learn the statistics of renewable generations and human behaviors. Nevertheless, existing learning-based charging strategies necessitate high-quality training data and may falter when the data distribution shifts over time. This highlights the concept of Out-of-Distribution (OOD), a critical issue in machine learning~\cite{hendrycks2016baseline,teney2020value} where models face data markedly different from their training sets. 

% In such circumstances, resulting in performance that falls short of traditional methods. 

\begin{figure}[ht]
\centering
\includegraphics[scale=0.2]{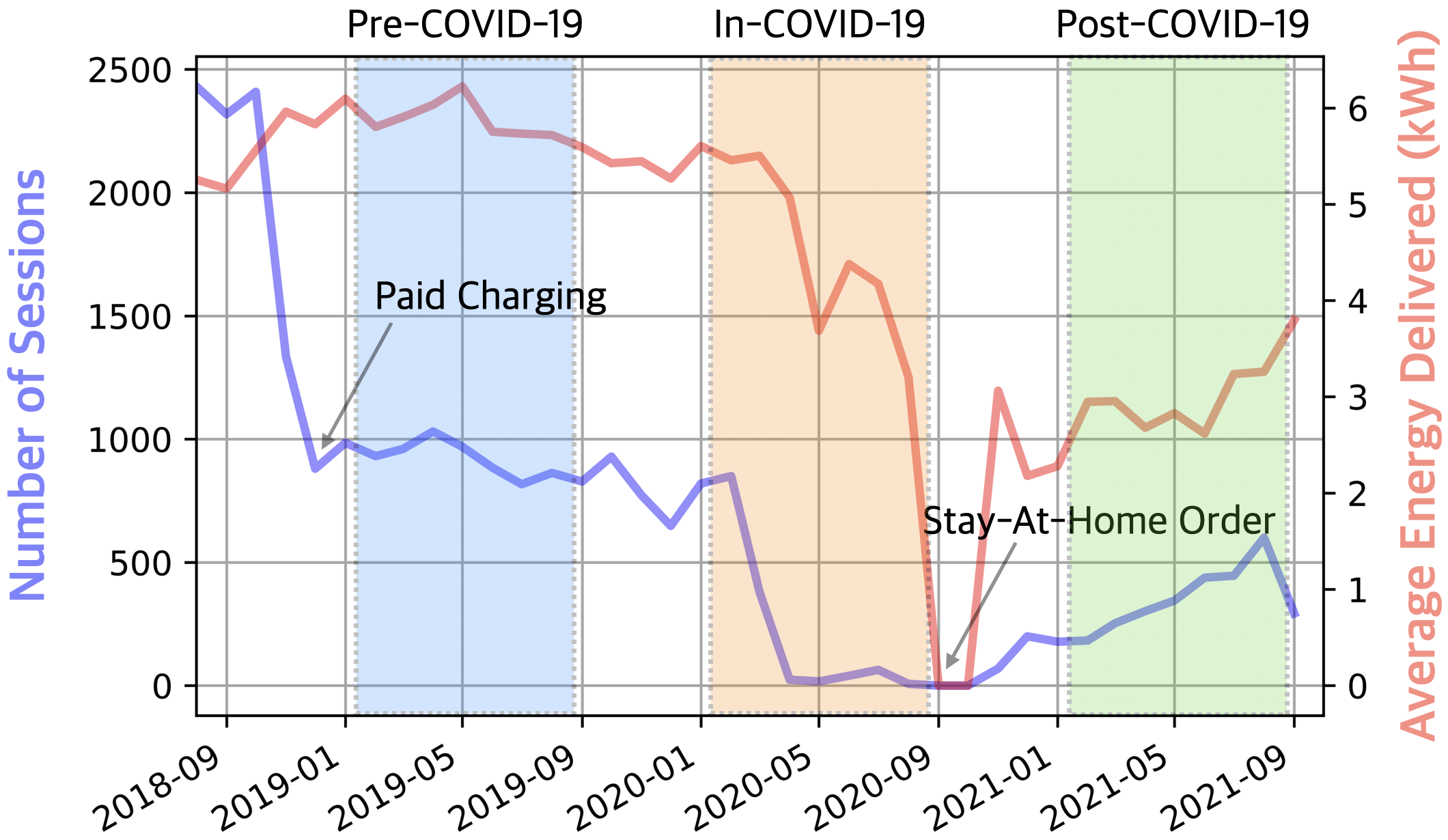}
\caption{Impact of pricing and social distancing policies on charging behaviors in the ACN-Data~\cite{lee2018large} {where ``ACN'' represents Adaptive Charging Network}. \textit{Blue Curve}: number of EV charging sessions; \textit{Red Curve}: average energy delivered to users.}
\label{fig:data_shift}
\end{figure}

\begin{figure*}[ht]
\centering
\includegraphics[scale=0.4]{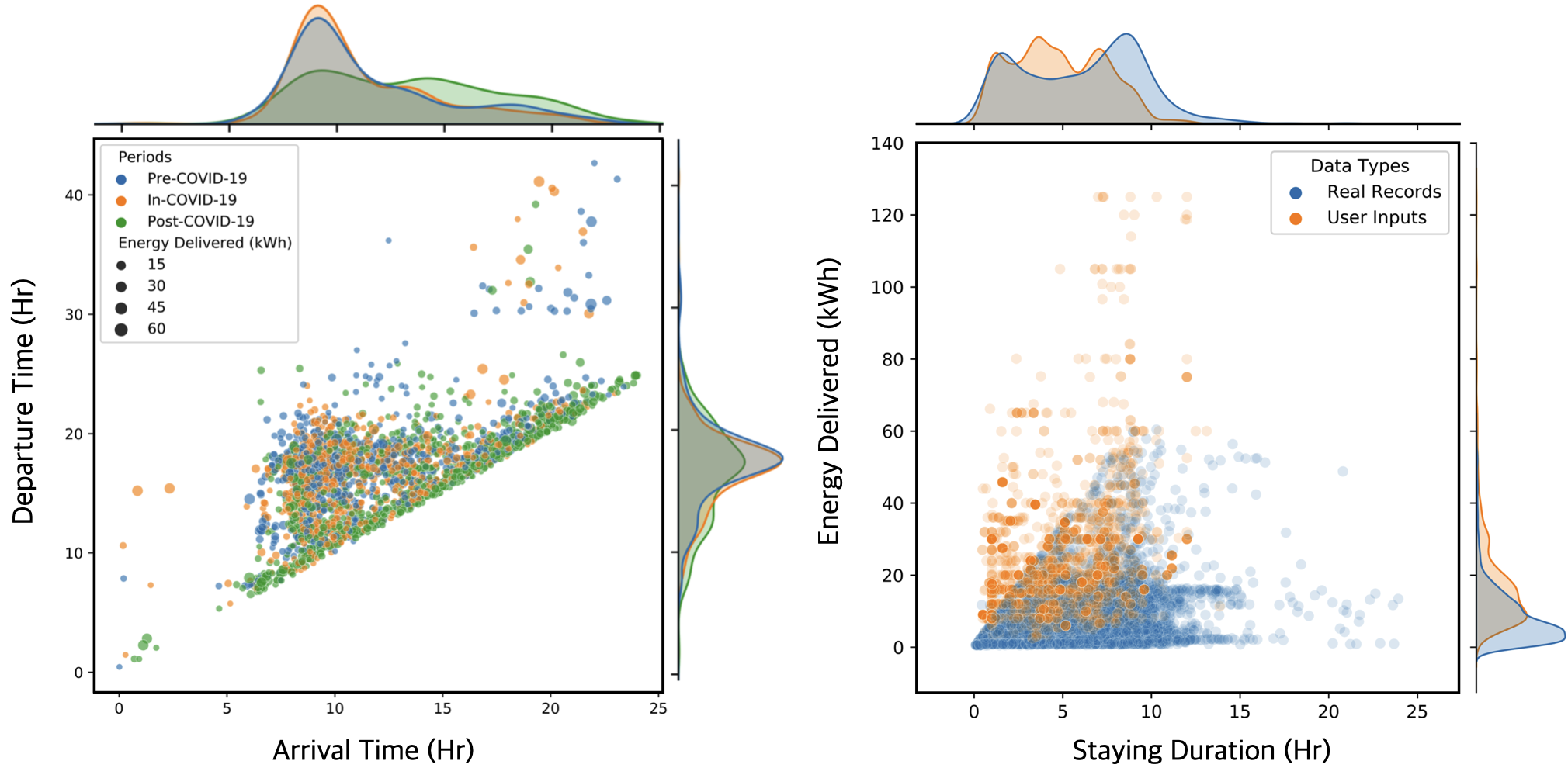}
\caption{\textsc{OOD phenomenon observed in ACN-Data~\cite{lee2019acn}}. (Left) \textit{Distribution Shift}:  Charging behaviors affected by the change of social distancing policies; (Right) \textit{User Input Data}: Error in user inputs in terms of the energy demand and departure time. The user input predictions are used in the practically implemented MPC at Caltech EV charging station. See Section~\ref{sec:mpc_baseline} for more details.}
\label{fig:compare_together}
\end{figure*}

Charging behaviors of EV drivers are often nonstationary and vulnerable to reward and incentive policies~\cite{li2021coordinating,kacperski2022impact}, planning of nearby charging stations~\cite{arias2017robust}, electricity pricing~\cite{lee2019acn}, curfew rules~\cite{lee2019acn}, and other factors. For instance, real user charging profiles recorded at the Caltech EV charging station, which comprise 54 chargers, with data collected from September 2018 to September 2021~\cite{lee2019acn} 
instantiate the phenomenon of OOD in real-world EV charging platforms. As shown in Figure~\ref{fig:data_shift}, the charging station switched from free charging to imposing a fee of \$0.12 / kWh in November 2018, and led to a decrease of active charging sessions. 
Furthermore, notable 
changes of charging behaviors can be witnessed in the repercussions of the COVID-19 pandemic. 
In early 2021, social distancing policies led to a significant reduction in both the total number of EV charging sessions and the overall energy demand, and a demand recovery can be observed in 2022. Simultaneously, as shown in the left column of Figure~\ref{fig:compare_together}, there was a significant shift in the data distribution of arrival times before and after the COVID-19 outbreak. 

It is important to highlight that classic learning-based charging algorithms, such as reinforcement learning (RL) trained using data from the pre-COVID era, are likely to experience a significant performance degradation when applied in the post-COVID period (see Section~\ref{sec:exp} for detailed numerical results). As a result, more robust control-based scheduling methods have been employed widely in real charging facilities. A representative example is the Model Predictive Control (MPC), with future departure times and battery capacities provided by user inputs through a mobile app, has been in use at the Caltech charging station since 2018~\cite{lee2021adaptive}. Notably, these user-input parameters diverge significantly from the ground truth, as shown in the right column of Figure~\ref{fig:compare_together}. This discrepancy arise due to multiple reasons, e.g., drivers acting as players in a game with unexpected Nash equilibria, a lack of user engagement with the mobile app, or simply a human error. Despite being sub-optimal, this approach has proven its robustness as detailed in~\cite{lee2021adaptive}, even for worst-case scenarios.

In summary, both the advanced learning-based  and control-based algorithms~\footnote{{It is worth highlighting that, as thoroughly evaluated in the Adaptive Charging Network (ACN) infrastructure at Caltech and JPL~\cite{lee2021acn,lee2021adaptive}, MPC-type scheduling outperforms other baseline methods including Round Robin, First Come First Serve (FCFS), Earliest Deadline First (EDF), Least Laxity First (LLF), closely approximating the offline optimal results.}} are not close to optimal for realistic EV charging tasks in the following sense -- (1). The learning-based algorithm achieves higher scheduling efficiency on average, but cannot guarantee the \textit{worst-case performance} due to distribution shifts; (2). The MPC scheduling is more robust against non-stationary environments, but is \textit{sub-optimal}.
Together, this leads us to a crucial question examined in this paper:

\vspace{3pt}

\textit{Can we design an OOD-aware EV charging algorithm, taking advantage of both learning-based and MPC scheduling?} 

% \textit{Can we take advantage of both ML policy and the classic MPC scheduling for more effective OOD EV charging?} 

\vspace{5pt}

\textbf{Main Contributions.}
Tackling this challenge, in this paper, we propose a novel ``learning-augmented'' algorithm that synergizes a value-based RL policy and the MPC scheduling. Our main contributions are {two-fold}.

\textit{A novel OOD-aware charging algorithm}. 
{ We first introduce an OOD-aware policy, denoted by \ouralg, which not only achieves near-optimal average performance under stable conditions but also ensures robust performance in the face of distribution shifts. The proposed \ouralg enhances the MPC scheduling currently used in real EV charging stations at Caltech and JPL garages. To the best of our knowledge, our work represents one of the first investigations into EV charging scheduling algorithms with OOD-data.} 

%Section~\ref{sec:policy} introduces

% This policy extends the \textsc{PROP} policy in~\cite{li2023beyond} for a single-trajectory Markov decision process (MDP), which requires the existence of a Wasserstein robust baseline (see Definition 3 in~\cite{li2023beyond}). 

% First, in Section~\ref{sec:policy}, we present a novel learning-augmented policy~\ouralg that achieves near-optimal average performance when no distribution shift is present, and ensure robust worst-case performance when a distribution shift occurs, for an EV charging problem with installed solar panels and batteries given in Section~\ref{sec:model}. This framework extends the \textsc{PROP} policy in~\cite{li2023beyond} for a single-trajectory Markov Decision Process (MDP), which requires the existce of a Wasserstein robust baseline (see Definition 3 in~\cite{li2023beyond}). In this work, we extend the applicability of the learning-augmented policy to the solar-powered scheduling problem of EVs. Note that verifying that the MPC scheduling indeed satisfies the Wasserstein robustness property is nontrivial, due to the non-linearity of the charging dynamics. Furthermore, the exact system state is not directly available at each time slot since the EV battery capacity is assumed unknown (this is a practical setting as in the Caltech Adaptive Charging Network (ACN)).

% results of \ouralg in Section~\ref{sec:main} 

%of solar-powered EV charging as detailed in Section~\ref{sec:model}

\textit{Theoretical guarantee}. {In addition, we offer theoretical results of \ouralg.} Aligning with traditional definitions of \textit{consistency} and \textit{robustness} for algorithms with predictions  (cf.~\cite{purohit2018improving}), we provide performance guarantees of \ouralg by characterizing its consistency and robustness tradeoff in Theorem~\ref{thm:dynamic_regret}. 
{Our approach broadens the application of learning-augmented policies to address the complexities of EV scheduling in a specific context of solar-powered EV charging}, {where satisfying the Wasserstein robustness property of the MPC scheduling is nontrivial due to nonlinear charging dynamics and the practical assumption of unknown EV battery capacities, as seen in the Caltech charging stations~\cite{lee2021acn}}.

% The \ouralg policy demonstrates $\left(1+O(\overline{W})\right)$-consistent and $\left(\frac{2\xi C^2(1 + C^2)(1 + \overline{A}^2 + \overline{B}^2)}{\mu (1 - \overline{\decayfactor})^2} +O(\overline{W})+o(1)\right)$-robust for some positive  $C$ and $\overline{\lambda}$, where $\xi,\mu,\overline{A},\overline{B},$ and $\overline{W}$ are constants that depend on the charging system dynamics. 

% Second, in Section~\ref{sec:main}, we provide a theoretical analysis of the \ouralg policy. Following the classic \textit{consistency} and \textit{robustness} definitions in the analysis of algorithms with predictions (cf.~\cite{purohit2018improving}). We bound the expected dynamic regret of the \ouralg policy in Theorem~\ref{thm:dynamic_regret}, based on which, the guarantee of \ouralg is summarized in Corollary~\ref{coro:algorithm_performance} that \ouralg is $\left(1+O(\overline{W})\right)$-consistent and $\left(\frac{2\xi C^2(1 + C^2)(1 + \overline{A}^2 + \overline{B}^2)}{\mu (1 - \overline{\decayfactor})^2} +O(\overline{W})+o(1)\right)$-robust for some positive  $C$ and $\overline{\lambda}$, where $\xi,\overline{A},\overline{B},$ and $\overline{W}$ are constants that depend on the charging system dynamics. 

% \textit{Numerical results.} Finally, in Section~\ref{sec:exp}, the efficacy of \ouralg is validated through experimental results for realistic OOD scenarios built upon the ACN-Data~\cite{lee2019acn}.

% Experimental results validates a near-
% optimal trade-off between consistency and robustness in OOD scenarios.

\textbf{Related Literature.}
{
The growing interests of investigating learning-based charging algorithms have led to
the emergence of a rich literature, characterized into three categories below.}

{
\subsubsection{Reinforcement learning for EV charging} The use of 
Reinforcement Learning (RL) has been explored in the domain of EV charging~\cite{qian2023impact}, as surveyed in~\cite{chen2022reinforcement}. For example, Wan \textit{et al.}~\cite{wan2018model} developed a deep RL framework that employs a representation network to capture features from electricity prices and a Q-network to estimate the optimal action-value function.} Similarly, Li \textit{et al.}~\cite{li2019constrained} tackled the constrained EV charging scheduling problem by framing it as a constrained Markov decision process, introducing a safe deep RL strategy that derives schedules under uncertainty. 
{Recently, the authors of~\cite{su2023electric} introduced a scalable, real-time EV charging guidance algorithm using Multi-agent Reinforcement Learning (MARL) to address driver inconvenience and mileage anxiety. Similarly, in~\cite{dong2023multi}, an MARL mechanism that autonomously schedules EV battery discharging to optimize peak shaving was proposed. The study in~\cite{yang2024multiagent} developed a multi-agent value-based pricing policy for fast charging stations in integrated transportation and power networks, reducing costs and balancing demand while improving traffic and energy efficiency. Going beyond the routine application of RL-related methods,~\cite{wang2023transfer} designed a deep transfer RL method for EV charging that adapts an existing RL-based strategy to new environments, reducing the need for extensive retraining and data collection. Despite recent advancements, applying RL to EV charging still faces challenges in OOD scenarios.
}

% Reinforcement Learning-based methods have attracted much attention due to their success in many domains~\cite{silver2016mastering}, and have been applied to EV charging scheduling recently~\cite{chen2022reinforcement}. 
% For example, Wan~\textit{et al.}~\cite{wan2018model} proposed a Deep Reinforcement Learning (DRL) based approach that extracts the features from electricity prices with a representation network and approximates the optimal action-value function with a Q-network. 
% Li~\textit{et al.}~\cite{li2019constrained} formulated the constrained EV charging scheuling problem as a constraint Markov Decision Process (CMDP), and proposed a safe DRL approach to determine the optimal schedules without the knowledge of randomness and the constraint. 
% Despite the fact that some scalable deep RL algorithms exist, applying them requires certain conditions and may not work for large-scale EV charging problems.

\subsubsection{Learning-augmented scheduling algorithms}

The integration of machine-learned insights into online algorithms, pioneered by Mahdian \textit{et al.}~\cite{mahdian2012online}, has revolutionized the design of such algorithms. Purohit \textit{et al.}~\cite{purohit2018improving} further advanced this field by introducing ``robustness'' and ``consistency'' within a formal framework based on competitive ratios. This paradigm has since been widely adopted, enhancing online algorithms with black-box advice across various domains~\cite{purohit2018improving,banerjee2020improving,rohatgi2020near,lykouris2021competitive,im2022parsimonious,antoniadis2020online,li2022robustness,anand2022online,christianson2022chasing,li2023beyond}. These approaches typically involve meta-strategies that meld online algorithms with predictive models, requiring a manually set trust hyper-parameter.
% More recent work by Li \textit{et al.}~\cite{li2023certifying} explored a balance between competitiveness and stability in nonlinear control settings, albeit limited to a linear quadratic regulator without a theoretical framework for trust parameter selection.
% Diakonikolas \textit{et al.}~\cite{diakonikolas2021learning} expanded on this by incorporating distributional advice into the black-box model. 
{ Recent works have focused on scheduling problems from an algorithmic point of view, as highlighted in key studies like~\cite{lindermayr2023speed,balkanski2024energy}.} Despite these developments, existing learning-augmented algorithms have yet to offer a solution that adequately addresses the unique challenges posed by solar-powered EV charging dynamics.

\subsubsection{{Integration of model-free and model-based methods}}

Our method situates within the broader framework of integrating model-free RL with model-based policies. The landscape of learning-based MPC has been rapidly evolving. A salient example is the terminal deep value MPC, which synergizes MPC with an ancillary neural network that encodes a terminal function, ensuring asymptotic stability~\cite{rosolia2017learning}. This approach has been successfully applied to autonomous robot control~\cite{karnchanachari2020practical,brunke2022safe} and operator-aggregator coordination~\cite{li2021learning}. Furthermore, MPC has been instrumental in facilitating safe exploration within RL paradigms~\cite{koller2018learning}. Hansen \textit{et al.} have innovated upon this by integrating temporal difference learning with MPC, consequently enhancing sample efficiency and asymptotic performance~\cite{hansen2022temporal}. {Additionally, control baselines have been strategically employed as a regularizer during RL's exploration phase to curb the inherent high variability~\cite{cheng2019control}, based on the Temporal Difference (TD)-error.} Despite these significant advances and the apparent benefits of melding model-free RL with control priors, a tailored application to EV charging systems—particularly in the face of pronounced distribution shifts as illustrated in Figure~\ref{fig:compare_together}—has not yet been realized.

{
\section*{Nomenclature Table}
	\addcontentsline{toc}{section}{Nomenclature}
	\subsection*{Key Symbols in EV Charging Model}
	\begin{IEEEdescription}[\IEEEusemathlabelsep\IEEEsetlabelwidth{$g_{\mathcal{S}},g_{\mathcal{A}}$}]
	\item[$\nt$] Length of the scheduling time
horizon
        \item[$\nm$] Number of chargers
	\item[$t$] Time index
	\item[$s_t$] System state in $\mathcal{S}$
	\item[$a_t$] Scheduling action $\mathcal{A}$
 	\item[$\widetilde{s}_t$] Estimated system state in $\mathcal{S}$
	\item[$c_t$] Charging cost
	\item[$\pi$] Scheduling policy
	\item[$P_t$] Charging transition probability
 	\item[$A_t,B_t$] Matrices of charging dynamics
  	\item[$Q_t,R_t$] Matrices of charging costs
\item[$g_{\mathcal{S}},g_{\mathcal{A}}$] State and action projection functions
\item[$\ell_t$] Perturbation from human charging behaviors
\item[$h_t$] Perturbation from solar injection
 \item[$\mathcal{D}$]  Collection of all charging sessions
\end{IEEEdescription}
\vspace{-5pt}
\subsection*{Key Symbols in OOD-Aware EV Charging Algorithm}
\begin{IEEEdescription}[\IEEEusemathlabelsep\IEEEsetlabelwidth{$u_t(j)$}]
	% \item[$Q_t$] Q-value function at time $t$
\item[$Q^{\star}_t,\widetilde{Q}_t$] Optimal and estimated Q-value function at time $t$
\item[$\overline{\pi}_t,\widetilde{\pi}_t$] MPC baseline and value-based neural policy
\item[$\varepsilon$] Error between Q-value functions
\item[$R_t$] Awareness radius
\item[$\mathsf{TD}_t$] Estimated  Temporal Difference (TD)-error
\end{IEEEdescription}
}

\vspace{6pt}

{
\textbf{Notation and Conventions. } 
Throughout this paper, we use $\mathbb{P}\left(\cdot\right)$ and $\mathbb{E}\left(\cdot \right)$ to denote the probability distribution and expectation of random variables. The symbol $\|\cdot\|$ denotes the $\ell_2$-norm for vectors and the matrix norm induced by the $\ell_2$-norm. We use a subscript $(\cdot)_t$ to represent a length-$k$ vector $s_t=(s_{t1},\ldots,s_{tk})$ at time $t$ whose $i$-th coordinate is written as $s_{ti}$.}
{
The rest of the paper is organized as follows. We present our OOD-aware EV charging problem in Section~\ref{sec:model}. In Section~\ref{sec:policy}, 
we introduce an OOD-aware EV charging policy, denoted by \ouralg, whose performance is theoretically analyzed in Section~\ref{sec:main}. Numerical results based on realistic EV charging scenarios are provided in Section~\ref{sec:exp}. Finally, we conclude this paper in Section~\ref{sec:conclusion}.
}

% Our method belongs to the general contexts that combine model-free RL and model-based policies.
% Recently, a series of learning-based MPC has been proposed. For example, terminal deep value MPC combines MPC and a separate neural network representing a terminal function that  guarantees asymptotic stability~\cite{rosolia2017learning}, autonomous robot control~\cite{karnchanachari2020practical,brunke2022safe}, and operator-aggregator coordination~\cite{li2021learning}. MPC has been used for safe exploration in RL~\cite{koller2018learning}. Hansen et al. combine temporal difference learning and MPC to improve sample efficiency and asymptotic performance~\cite{hansen2022temporal}. Control baselines have been used as a regularizer in the exploration stage to mitigate the high variability issues in RL~\cite{cheng2019control}. Despite the huge potential benefits of combining model-free RL and control priors as revealed in the existing approaches, none of existing results can be directly applied to EV charging, especially when significant distribution shifts present, for instance, the scenario depicted in Figure~\ref{fig:compare_together}.

% Neural network
% dynamics for model-based deep reinforcement learning with model-free fine-tuning

% Dual Stochastic MPC for Systems with Parametric and Structural Uncertainty.

% Regret Bounds for Adaptive Nonlinear Control

\section{Problem Formulation}
\label{sec:model}
% In this section, we define a mathematical formulation of the \textit{adaptive EV charging scheduling problem}. 

\subsection{Preliminary: Markov Decision Process}
\label{sec:mdp}
The general EV charging scheduling problem considered in this work is formulated as a finite Markov Decision Process (MDP). 
Denote by $\nt>0$ the length of the scheduling time horizon and write $[\nt]\coloneqq \{0,\ldots,\nt-1\}$. Let $s_t\in \mathcal{S}$  and $a_t\in \mathcal{A}$ be the system \textit{state} and \textit{scheduling action} at time $t\in [\nt]$ where $\mathcal{S}\subseteq\mathbb{R}^{\nn}$ and $\mathcal{A}\subseteq\mathbb{R}^{\nm}$ are safety state and action spaces defined by operational constraints. Let $P_t:\mathcal{S}\times\mathcal{A}\rightarrow \mathcal{P}_{\mathcal{S}}$ be the charging transition probability, where $\mathcal{P}_{\mathcal{S}}$ is a set of probability measures on $\mathcal{S}$. Given a charging action $a_t$ at time $t\in [\nt]$, the EV state evolves as
$s_{t+1} \sim P_t(\cdot|s_t,a_t), \ t\in [\nt]$. We denote a scheduling policy by $\pi\coloneqq (\pi_t:t\in [\nt])$ where each $\pi_t:\mathcal{S}\rightarrow\mathcal{A}$ decides a charging action $a_t$, after observing the EV state $s_t$ at time $t\in [\nt]$. 
We consider a time-varying charging cost (or negative reward) $c_t:\mathcal{S}\times\mathcal{A}\rightarrow \mathbb{R}_+$ with variability induced by solar generation and electricity costs. Overall, our goal is to find a policy $\pi$ that minimizes the following expected costs:
\begin{align}
\label{eq:reward}
    J(\pi) \coloneqq & \mathbb{E}\left[\sum_{t\in [\nt]} c_t\left(s_t,\pi_t(s_t)\right)\right],
\end{align}
where the randomness in the expectation $\mathbb{E}(\cdot)$ is from the transition dynamics $\mathbb{P}=(P_t:t\in [\nt])$. This MDP charging model is compactly denoted by $\mathsf{MDP}(\mathcal{S},\mathcal{A},\nt,\mathbb{P},c)$.

\subsection{OOD-Aware EV Charging Problem}
\label{eq:LQ-ACN}

Next, we introduce a linearized solar-powered EV charging model, which is a special case of the finite-horizon MDP in Section~\ref{sec:mdp}. It is worth emphasizing that this concrete charging model serves as a practical and illustrative example. Our proposed method remains applicable for the general MDP model without loss of generality. 
Consider a problem of charging a fleet of EVs at a charging station equipped with solar panels (see Figure~\ref{fig:model}). Let $\nm$ be the number of EV chargers.  At each time $t\in [\nt]$, new EVs may arrive and  EVs being charged may depart from previously occupied chargers. Each new EV $j$ induces a charging session, which can be represented by a tuple $\left(\alpha_j,\delta_j,\kappa_j, i\right)$ where at time $\alpha_j$, EV $j$ arrives at charger $i\in \{1,\ldots,\nm\}$, with an EV battery capacity $\kappa_j$, and depart at time $\delta_j$. Denote by $\mathcal{D}\coloneqq \left\{\left(\alpha_j,\delta_j,\kappa_j, i\right): j=1,\ldots,k\right\}$ the collection of all charging sessions.\footnote{Note that $\mathcal{D}$ is stochastic and nonstationary, due to the OOD issues. In the following context, for the ease of presentation, we define our charging model assuming $\mathcal{D}$ is given.
}

\subsubsection{Charging states and actions}
Denote by $e_t=(e_{ti}:i=1,\ldots,\nm)\in\mathbb{R}_+^\nm$ the State-of-Charge (SoC) of the $\nm$ chargers, i.e., $\smash{e_{ti}>0}$ if an EV is charging at charger-$i$ and $\smash{e_{ti}}$ (kWh) is the amount of energy that needs to be delivered; $\smash{e_{ti}<0}$ if the installed battery is charged; otherwise   $\smash{e_{ti}=0}$. Let $b_t=(b_{ti}:i=1,\ldots,\nm)\in\mathbb{R}_+^\nm$ be the allocation of energy to the $\nm$ chargers at time $t\in [\nt]$. There is a charging limit $\overline{b} >0$ for each charger so that $\smash{b_{ti}\leq \overline{b} }$ for all $i$ and $t$. The charging rates also need to satisfy {an energy constraint such that $\|b_t\|_{1}\leq \gamma$.} We unify this EV charging mode by defining a state $s_t=(e_t\| b_t)\in \mathbb{R}^{\nn}$ where $b_t\in\mathbb{R}^{\nm}$ contains the charging rates of all $\nm$ chargers, with $\nn=2\nm$. The symbol $(\cdot\|\cdot)$ concatenates two vectors. Define an action vector $a_t\in\mathbb{R}^{\nm}$ with each $a_{ti}\in [\underline{a},\overline{a}]$ that controls the change of charging rates from the connected grid.  The state and action spaces, denoted by $\mathcal{S}$ and $\mathcal{A}$ define constraints on the battery capacities and charging rates. In our context, they are instantiated as
\begin{align}
\label{eq:state_space}
\mathcal{S} & \coloneqq \left\{(e\| b)\in \mathbb{R}_+^{\nn}: \|b\|_{1}\leq \gamma, b_i\leq \overline{b}, \forall i \right\},\\
\label{eq:action_space}
\mathcal{A} & \coloneqq \left\{a\in \mathbb{R}^{\nm}: \underline{a}\leq a_i\leq \overline{a}, \forall i \right\}.
\end{align}

\subsubsection{Charging dynamics}
The charging state updates according to the battery dynamics with unknown perturbations occur since the plugging and unplugging of EV chargers and solar injections, therefore are vulnerable to OOD situations as demonstrated in Figure~\ref{fig:data_shift} and~\ref{fig:compare_together}. Below we introduce a formulation of the OOD-aware EV charging problem, based on the following charging dynamics as a special case of $\mathsf{MDP}(\mathcal{S},\mathcal{A},\nt,\mathbb{P},c)$ defined in Section~\ref{sec:mdp}:
\begin{align}
\label{eq:original_dynamics}
     s_{t+1}=& g_{\mathcal{S}}\Big [\underbrace{A_ts_t+B_t g_{\mathcal{A}}(a_t)}_{\textit{Battery Dynamics}} + \underbrace{\ell'_t-\Delta h'_t}_{\textit{Behavior/Solar Perturbations}}\Big ],
\end{align}
where $g_{\mathcal{S}}:\mathbb{R}^{\nn}\rightarrow\mathcal{S}$ is a safety function defined as
\begin{align*}
g_{\mathcal{S}}(s)_i\coloneqq \begin{cases}
0 & \ \text{if } t<\delta_j\leq t+1 \text{ and } (\delta_j,i)\in \mathcal{D}, \\
\mathrm{Proj}_{\mathcal{S}}(s)_i & \ \text{otherwise}.
\end{cases}.
\end{align*}
Here, we slightly abuse the notation and write $(\delta_j,i)\in \mathcal{D}$ when  $(\delta_j,i)$ appears in a charging session tuple; $\mathrm{Proj}_{\mathcal{S}}(s)=\min_{x\in\mathcal{S}} \|x-s\|_2$ projects states to $\mathcal{S}$ (see~\eqref{eq:state_space}). The event $t<\delta_j\leq t+1$ implies that the active charging session ends and the EV departs, so the $i$-th entry $g_{\mathcal{S}}(s)_i$ is reset to $0$. Similarly,
$
g_{\mathcal{A}}(a)\coloneqq \mathrm{Proj}_{\mathcal{A}}(a).
$

Moreover, block matrices $(A_t:t\in [\nt])$ and $(B_t:t\in [\nt])$ are system matrices modeling the time-varying behaviors of a non-ideal battery with the following forms:
\begin{align}
\label{eq:A_and_B}
A_t &=\begin{pmatrix}
 \bigone_{\nm}
  & \rvline & -\Delta\mu_t\bigone_{\nm} \\
\hline
  \bigzero & \rvline &
 \bigone_{\nm}
\end{pmatrix}, B_t=\begin{pmatrix}
 0 \\
\hline
 \beta_t \bigone_{\nm}
\end{pmatrix}, \\ 
\nonumber
\ell'_t &=\begin{pmatrix}
\ell_t \\
\hline
\bigzero_{\nm}
\end{pmatrix}, h'_t=\begin{pmatrix}
\bigzero_{\nm}\\
\hline
h_t 
\end{pmatrix},
\end{align}
{where $(\ell_t\in\mathbb{R}^{\nm}:t\in [\nt])$ are stochastic perturbations caused by human charging behaviors so that $\ell_{ti}=\kappa_j$ when $\left(\alpha_j,\delta_j,\kappa_j, i\right)$ is a charging session with $t<\alpha_j\leq t+1$ (a new charging session starts), and $h_{ti}$ is the solar injection at the charger $i$ at time $t\in [\nt]$.
In~\eqref{eq:original_dynamics} and~\eqref{eq:A_and_B},
the parameter $\Delta$ is the time span between two charging controls, $\mu_t\in [0,1]$ and $\beta_t\in [0,1]$ are charging efficiency and control efficiency factors. At each time $t\in [\nt]$, in the charging dynamics~\eqref{eq:original_dynamics}, the future perturbations $(\ell_\tau-\Delta h'_\tau:\tau\geq t)$ are unknown. The previous solar generations $(h_\tau:\tau<t)$ are observed, and the parameters $(A_t,B_t:t\in [\nt])$, $g_{\mathcal{S}}(\cdot)$ and $g_{\mathcal{A}}(\cdot)$ of the dynamical system are given beforehand.}

\subsubsection{Charging costs}
We consider quadratic charging costs given by
\begin{align}
\label{eq:quadractic_cost}
    c_t(s_t, a_t) = \frac{1}{2}\left( (s_t)^\top Q_t s_t + (a_t)^\top R_t a_t \right)
\end{align}
for every time step $t\in[\nt]$, as a special instance of the cost function in the general MDP model in Section~\ref{sec:mdp}. Here, $Q_t\in\mathbb{R}^{\nn\times\nn}$ and $R_t\in\mathbb{R}^{\nm\times\nm}$  are time-varying positive-definite matrices. This total cost penalizes the excessive use of installed batteries and incentivizes the utilization of solar generation.

% We make some standard assumptions, following those in the literature of online control \cite{lin2021perturbation,zhang2021regret,lin2022bounded}.
% The first assumption is that the cost functions are well-conditioned and the dynamical matrices are uniformly bounded.
% \begin{assumption}\label{assump:bounded-costs-and-dynamics}

\subsubsection{Model assumptions}
We adopt several standard assumptions, aligning with those prevalent in the literature of online control \cite{lin2021perturbation,zhang2021regret,lin2022bounded}. 
The first assumption requires the well-conditioning of costs and dynamical matrices.

\begin{assumption}\label{assump:bounded-costs-and-dynamics}
Let $\overline{A}, \overline{B}$, $\overline{W}$, $\mu$, and $\xi$ be positive constants.
For any $t \in [\nt]$, the matrices $A_t,B_t,Q_t,R_t$ and the perturbations satisfy $\norm{A_t} \leq \overline{A}, \norm{B_t} \leq \overline{B}, \norm{\ell'_t-\Delta h'_t} \leq \overline{W}$, and
\[\mu I_n \preceq Q_t \preceq \xi I_n, \ \ \mu I_m \preceq R_t \preceq \xi I_m, \ \ \mu I_n \preceq P \preceq \xi I_n.\]
Furthermore, let $J^{\star}\coloneqq \inf_{\pi}J(\pi)$ denote the optimal expected cost. We assume $J^\star=\Omega(\nt)$.
\end{assumption}

The next assumption guarantees that for arbitrary bounded perturbation sequences $(\ell'_t-\Delta h'_t:t\in [\nt])$, there exists a controller that can stabilize the charging state.

\begin{assumption}\label{assump:uniform-stability}
For any $t, t' \in [\nt], t \leq t'$, define a block matrix
\begin{align*}
\Phi_{t, t'}^t \coloneqq \begin{bmatrix}
    I & & & & \\
    - A_t & - B_t & I & & \\
     & & \ddots & &   \\
     & & - A_{t'-1} & - B_{t'-1} & I
    \end{bmatrix}.
\end{align*}
We assume $\sigma_{\min}\left(\Phi_{t, t'}\right) \geq \sigma$ for some positive constant $\sigma$, where $\sigma_{\min}(\cdot)$ denotes the smallest singular value of a matrix.
\end{assumption}

In particular, Assumption~\ref{assump:uniform-stability} holds provided that there is an exponentially stabilizing policy for the dynamical system in~\eqref{eq:original_dynamics}, \eqref{eq:A_and_B}, and~\eqref{eq:quadractic_cost} (see~\cite{li2023beyond,lin2021perturbation} for a proof).

\subsection{Scheduling Baseline: Model Predictive Control}
\label{sec:mpc_baseline}
Prior to delving into our proposed policy, we highlight the Model Predictive Control (MPC) as a foundational baseline approach to scheduling EV charging, as detailed in~\cite{lee2021adaptive}. 

Let $(\widetilde{\ell}_{\tau|t}:\tau\in [t:t'-1])$ be a sequence of predictions of the stochastic charging behavioral perturbations $({\ell}_{\tau|t}:\tau\in [t:t'-1])$ received at time $t\in [\nt]$ with a prediction horizon $k$ and $t'\coloneqq\min\{t + k, \nt-1\}$ with a prediction horizon $k$.\footnote{For instance, the MPC implemented at Caltech EV charging station sets each $\widetilde{\ell}_{\tau|t}$ as user-input predictions (shown at the right of Figure~\ref{fig:compare_together}), generated from user-provided session information $(\alpha_j,\widetilde{\delta}_j,\widetilde{\kappa}_j, i)$ for all arrived sessions $j\in \{j:t\geq \alpha_j\}$.}
% We assume that the user-input predictions satisfy the following assumption.
% \begin{assumption}
% The user-input predictions of their session information $\widetilde{c}_j=(\alpha_j,\widetilde{\delta}_j,\widetilde{\kappa}_j, i)$ satisfy that $\left\|w_t-\widetilde{w}_t\right\|\leq \rho$ for all $t\in [\nt]$ for some constant $\rho>0$.
% \end{assumption}
Furthermore, write $(\widetilde{h}_{\tau|t}:\tau\in [t:t'-1])$ the predictions of solar generations received at time $t$.
At each time $t\in [\nt]$, define $\widetilde{w}_{t}\coloneqq\widetilde{\ell}'_{t}-\Delta \widetilde{h}'_{t}$. 
The MPC estimates the current state at time $t$ as
\begin{align}
\nonumber
\widetilde{s}_t=& \widetilde{g}_{\mathcal{S}}\Big(A_{t-1}s_{t-1}+B_{t-1}g_{\mathcal{A}}(a_{t-1}) + \widetilde{w}_{t-1}\Big)\\
\label{eq:charging_state_estimate}
     \coloneqq & g_{\textsf{MPC}}\left(s_{t-1},a_{t-1}, \widetilde{w}_{t-1}\right),
\end{align}
where the input departure time $\widetilde{\delta}_j$ and battery capacity $\widetilde{\kappa}_j$ can be different from the ground truth values in the charging session tuple $\left(\alpha_j,\delta_j,\kappa_j, i\right)$, and  $\widetilde{g}_{\mathcal{S}}$ is an estimate of the safety projection function ${g}_{\mathcal{S}}$ using the input departure times.

Given the estimated current system state $\widetilde{s}_t$, the predictions of future perturbations $(\widetilde{w}_{\tau|t}:\tau\in [t:t'-1])$  from time $t$ to $t'$ (assuming $t'>t$), and a terminal cost matrix $P_{t'}$, we consider the following MPC optimization:
\begin{subequations}
\label{eq:mpc}
\begin{align}
\min_{a_{t:(t'-1)\mid t}} &\sum_{\tau = t}^{t' - 1} c_\tau\left(s_{\tau\mid t}, a_{\tau\mid t}\right) + \frac{1}{2} s_{t'\mid t}^\top P_{t'} s_{t'\mid t}\\
\text{s.t. } s_{\tau+1\mid t} =& g_{\textsf{MPC}}\left(s_{\tau|t},a_{\tau|t},\widetilde{w}_{\tau|t}\right),  \tau \in [t:t'-1]\\
s_{t\mid t} =& \widetilde{s}_t.
\end{align}
\end{subequations}

It is worth highlighting that this MPC scheduling policy has been practically implemented at the Caltech EVSE~\cite{lee2021adaptive}. The term $s_{t'\mid t}^\top P_{t'} s_{t'\mid t}/2$ is a customized terminal cost that regularizes the last predictive state. 
With this notation, we define the per-step MPC update with user-input predictions formally in Procedure~\ref{alg:mpc_update}. At each time step, it solves a $k$-step predictive finite-time optimization and commits the first control action in the optimal solution with $k=\max_{j:t\geq \alpha_j}{\widetilde{\delta}_j}$. The terminal cost matrix $P_t$ is pre-determined. We present the update rule of this MPC baseline in Procedure~\ref{alg:mpc_update}, and denote the corresponding policy by $\pi_{\MPC}\coloneqq (\overline{\pi}_t:t\in [\nt])$.

\setcounter{algorithm}{0}
\begin{myprocedure}[t]
\SetAlgoLined
\SetKwInOut{Input}{Initialize}
\Input{Prediction horizon $k$}

% \textbf{Initialize:} {Prediction horizon $k$}

\BlankLine
{
Set $t'\gets \min\{t + k, \nt-1\}$

Receive $\left(A_{\tau}, B_{\tau}, Q_{\tau}, R_{\tau}, \widetilde{w}_{\tau}:  \tau\in [t:t'-1]\right)$

%Let $P \gets P_{h'}$ if $h + k \leq \nh-1$ and $P \gets Q_{\nh-1}^t$ otherwise

%{\color{gray}  \textit{// Use the default terminal cost if we cannot predict until the end of the episode.}}
% (\widetilde{s}_t)= \psi_{t, t'}(\widetilde{s}_t)[a_{t\mid t}]

Update policy $\overline{\pi}_{t}$ via~\eqref{eq:mpc}
}

\caption{\textsc{MPC-Update} (time $t$)}
\label{alg:mpc_update}
\end{myprocedure}

\begin{figure*}[ht]
\centering
\includegraphics[scale=0.32]{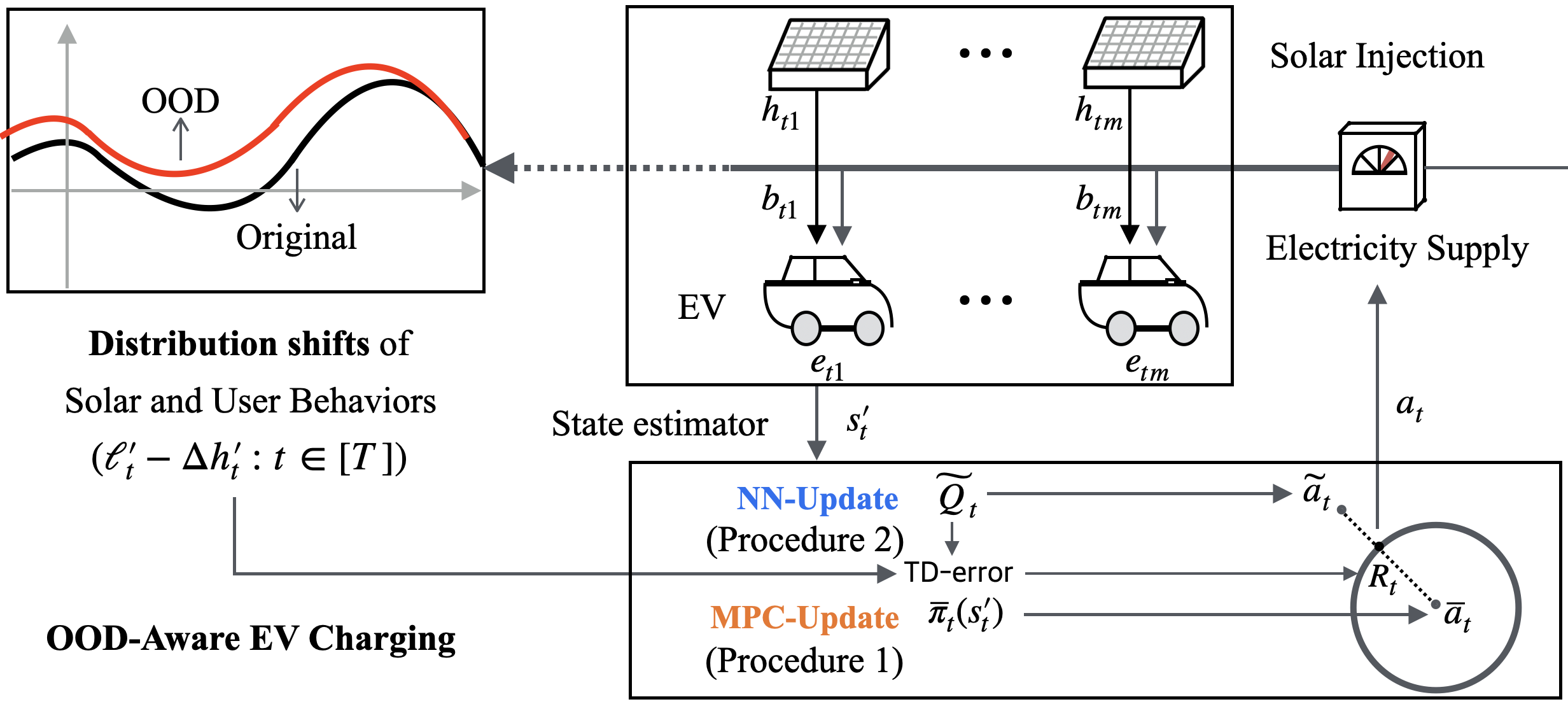}
\caption{{ A system diagram of \ouralg (see Algorithm~\ref{alg:ppp}), implemented at a workplace charging station with rooftop solar PV integration, as a motivating example of the OOD EV charging model (Section~\ref{sec:model}). The distribution of $(\ell'_t-\Delta h'_t:t\in [\nt])$ shifts when charging behaviors and solar injections change over time.}}
\label{fig:model}
\end{figure*}

\section{An OOD-Aware EV Charging Algorithm}
\label{sec:policy}

Classic learning-based scheduling algorithms~\cite{wan2018model,chen2022reinforcement,li2019constrained} may encounter high variability issues due to the changes of environments. In detail, for the EV charging problem described in Section~\ref{sec:model}, OOD of the perturbations $(\ell'_t-\Delta h'_t:t\in [\nt])$ will deteriorate the efficiency of a scheduling policy modeled by a Neural Network (NN). In this section, we present a novel scheduling algorithm that projects charging actions suggested by a neural policy $\pi_{\NN}=(\widetilde{\pi}_t:t\in [\nt])$ to a scheduling baseline $\pi_{\MPC}$ with a projection radius depending on the OOD level. Thus, the algorithm is OOD-aware, as illustrated in Figure~\ref{fig:model}.

\subsection{Value-Based Neural Policy}

We focus on value-based neural policy $\pi_{\NN}$ with a sequence of Q-value functions $(\widetilde{Q}_t:t\in [\nt])$ in a given episode. We define an error metric $\varepsilon$ in the $\ell_\infty$ norm that captures the discrepancy between the trained value functions and the optimal Q-value functions as
\begin{equation}
\label{eq:def_epsilon}
 \varepsilon \coloneqq \frac{1}{\nt}\sum_{t\in [\nt]}\Big\|\widetilde{Q}_{t}-Q_{t}^{\star}\Big\|_{\infty},
\end{equation} 
where $Q_t^{\star}$ denotes the optimal Q-value function defined as
\begin{align*}
Q_t^{\star}(s, a):=\inf _\pi \mathbb{E}_{P, \pi}\left[\sum_{\tau=t}^{T-1} c_\tau\left(s_\tau, a_\tau\right) \mid s_t=s, a_t=a\right]
\end{align*}
that satisfies the Bellman optimality equation.

\begin{myprocedure}[t]
\SetAlgoLined
\SetKwInOut{Input}{Initialize}
\Input{hyper-parameter $\phi_{\operatorname{old}}$ and learning rate $\gamma$ (detailed in Table~\ref{table:ddpg}, Appendix~\ref{table:ddpg})}

% \textbf{Initialize:} Actor and critic networks hyper-parameters $\phi_{\operatorname{old}}$ and $\theta_{\operatorname{old}}$, and learning rate $\gamma$ (detailed in Table~\ref{table:ddpg})

\BlankLine

{

Sample a batch $B$ from $\mathcal{D}$

Update estimated Q value-function by gradient descent with trajectory records $h\coloneqq \left(s, a, s', c\right)$ in $B$
 and the gradient
$$\nabla_\phi \frac{1}{|B|} \sum_{h \in B}\left(\widetilde{Q}_\phi(s, a)+c+Q_{\phi}(s',\widetilde{\pi}_{\theta}(s'))\right)^2
$$

%Let $P \gets P_{h'}$ if $h + k \leq \nh-1$ and $P \gets Q_{\nh-1}^t$ otherwise

%{\color{gray}  \textit{// Use the default terminal cost if we cannot predict until the end of the episode.}}

% Update policy by gradient descent with $\nabla_\theta \frac{1}{|B|} \sum_{s \in B} \widetilde{Q}_\phi\left(s, \mu_\theta(s)\right)$

Update hyper-parameters with
$
\begin{aligned}
\phi_{\operatorname{old}} & \leftarrow \gamma\phi+(1-\gamma) \phi_{\operatorname{old}} 
% \\
% \theta_{\operatorname{old}} & \leftarrow \gamma \theta+(1-\gamma) \theta_{\operatorname{old}}
\end{aligned}
$
}
\caption{\textsc{NN-Update} (time $t$)}
\label{alg:nn_update}
\end{myprocedure}

The following assumption of Lipschitz continuous value functions is standard~\cite{tang2020off,ghosh2022provably}.
\begin{assumption}
\label{ass:Q-Lip}
The machine-learned untrusted policy $\widetilde{\pi}$ is value-based.
The {Q}-value advice $\widetilde{Q}_{t}:\mathcal{S}\times \mathcal{A}\rightarrow\mathbb{R}$ is Lipschitz continuous with respect to $a\in\mathcal{A}$ for any $s\in\mathcal{S}$, with a Lipschitz constant $L_Q$ for all $t\in [\nt]$ and $\widetilde{Q}_t(s, a)-Q_t^{\star}(s, a)=o(T)$ for all $(s, a) \in \mathcal{S} \times \mathcal{A}$ and $t \in[\nt]$. 
\end{assumption}

We refer to the process of generating an estimated $Q$-value function the \textsc{NN-update} procedure.
In Procedure~\ref{alg:nn_update}, using the Deep Deterministic Policy Gradient (DDPG), we exemplify an approach of this procedure, which updates hyper-parameters of the neural policy and Q-value functions and provides an estimated $\widetilde{Q}_t$ (parameterized by $\phi$) at each time $t$. It is worth mentioning that our method is not limited to the specific policy update rule in Procedure~\ref{alg:nn_update}, and any RL algorithms that update value functions can be used as an \textsc{NN-update} procedure.

A key feature of the scheduling problem considered, compared with existing learning-augmented settings with state feedback~\cite{lin2021perturbation,li2023beyond} is that the exact state is unknown, therefore an estimated state $\widetilde{s}_t$ computed from~\eqref{eq:charging_state_estimate} is used as inputs for our scheduling policy described in the following section.

\subsection{Scheduling Policy Description}

Define $\mathcal{H}_t\coloneqq\left(A_{t}, B_{t}, Q_{t}, R_{t}, \widetilde{w}_{t}\right)$ for $t\in [\nt]$.
Our policy, \ouralg, outlined in Algorithm~\ref{alg:ppp} extends the \textsc{PROP} algorithm in~\cite{li2023beyond} to the specific domain of EV charging and generalizes to the setting of unknown state $s_t$ and general user-input predictions $(w_t:t\in [\nt])$ where $w_t=\ell'_t-\Delta h'_t$. 
The \ouralg policy is learning-based and stores the previous charging trajectories in a replay buffer $\mathcal{D}$. At each time $t$, user-input predictions, summarized as $\mathcal{H}_t$ are collected. The MPC baseline $\overline{\pi}_t$ and the value-based neural policy $\widetilde{\pi}_t$ are updated by implementing the procedures $\textsc{MPC-Update}$ (Procedure~\ref{alg:mpc_update}) and $\textsc{NN-Update}$ (Procedure~\ref{alg:nn_update}). Baseline and machine-learned actions $\overline{a}_t$, $\widetilde{a}_t$ are obtained by implementing MPC and minimizing the Q-value function $\widetilde{Q}_t$. Then, we compute an \textit{awareness radius}:
\begin{align}
\label{eq:budget}
R_t\coloneqq\Bigg[\left\|\widetilde{a}_t-\overline{a}_t\right\|_{2}-\beta\sum_{\tau=1}^{t}{\mathsf{TD}_{\tau}\left(\widetilde{s}_{\tau},\widetilde{s}_{\tau-1},a_{\tau-1}\right)}\Bigg]^+,
\end{align}
where we estimate the Temporal Difference (TD)-error based on previous trajectories in the replay buffer $\mathcal{D}$:
\begin{align}
\label{eq:td_error_approximate}
\mathsf{TD}_t\left(s_{t},s_{t-1},a_{t-1}\right) \coloneqq & c_{t-1} + \max_{a'\in\mathcal{A}}\widetilde{Q}_{t}\left(s_{t},a'\right)
- \widetilde{Q}_{t-1}.
\end{align}
{In~\eqref{eq:td_error_approximate}, $\beta$ is a tuning parameter that adjusts the impact of the estimated TD-error.}
%\left(s_{t-1},a_{t-1}\right)
Next, we project the machine-learned action $\widetilde{a}_t$ onto the ball with a radius $R_t$, centered at the baseline action $\overline{a}_t$ and obtain a new scheduling action $a_t$ such that $a_{t}=\mathrm{Proj}_{\mathcal{B}(\overline{a}_t,R_t)}\left( \widetilde{a}_t\right)$, as 
\begin{equation}
\nonumber
\mathrm{Proj}_{\mathcal{B}(\overline{a}_t,R_t)}(a) \coloneqq \argmin_{a'\in\mathcal{A}} \|a-a'\|_{2} \ \text{s.t. } \left\|a'-\overline{a}_t\right\|_{2}\leq R_t,
\end{equation}
with $\mathcal{B}(\overline{a}_t,R_t)\coloneqq \left\{a\in\mathcal{A}:\left\|a-\overline{a}_t\right\|_{2}\leq R_t\right\}$ defines an Euclidean ball centered at $\overline{a}_t$.
Finally, the charging environment is updated by scheduling according to $a_t$. Note that the \ouralg policy can be called repeatedly in an episodic setting so to update the value-based deep policy $\pi_{\NN}$.  The \ouralg described above is summarized in  Algorithm~\ref{alg:ppp}.

\begin{algorithm}[t]
\SetAlgoLined
\SetKwInOut{Input}{Initialize}\SetKwInOut{Output}{output}

\Input{Initial state $s_0$, parameters of $\pi_{\NN}=\left(\widetilde{\pi}_t:t\in [\nt]\right)$ and $\pi_{\MPC}=\left(\overline{\pi}_t:t\in [\nt]\right)$, replay buffer $\mathcal{D}$}
\BlankLine
\For{$t=0,\ldots,\nt-1$}{

Receive system and user-input predictions $\mathcal{H}_t$

{\it \color{gray}// Update the MPC scheduling baseline policy}

Implement \textbf{Procedure 1} 
$\textsc{MPC-Update}(\mathcal{H}_t)$, obtain $\overline{\pi}_{t}$

{\it \color{gray}// Update the learning-based policy}

Implement \textbf{Procedure 2} $\textsc{NN-Update}(\mathcal{D})$, obtain $\widetilde{Q}_t$

Set $\widetilde{\pi}(s)\in \argmin\widetilde{Q}_t(\cdot,s)$

$\overline{a}_t\leftarrow \overline{\pi}_{t}(\widetilde{s}_t)$

$\widetilde{a}_t\leftarrow\widetilde{\pi}_{t}(\widetilde{s}_t)$

Obtain $R_t$ via~\eqref{eq:budget}

Set action $a_{t}=\mathrm{Proj}_{\overline{\mathcal{A}}_t}\left( \widetilde{a}_t\right)$

Sample next state $s_{t+1} \sim P_t\left(\cdot|s_t,a_t\right)$

Estimate $\widetilde{s}_t$ via~\eqref{eq:charging_state_estimate}

Store $\left(\widetilde{s}_t,a_t,\widetilde{s}_{t+1},c_t\right)$ in replay buffer $\mathcal{D}$
}

\caption{OOD-Aware EV Charging (\ouralg)}
\label{alg:ppp}
\end{algorithm}

% In the next section, we provide theoretical guarantees for the \ouralg policy.

% , based on the ratio of expectations defined above

\section{Theoretical Guarantees}
\label{sec:main}

% \textit{Classical MPC controller.} Note that this is the algorithm implemented in the Caltech ACN. $\pi_{\text{MPC}}(s):\mathcal{S}\rightarrow\mathcal{A}$ defined as
% \begin{align*}
% \pi_{\text{MPC}}(s_t) = &\argmax_{a_t,\ldots,a_{t+w-1}} \sum_{\tau=t}^{t+w-1} \widehat{r}_\tau(s_\tau,a_\tau)\\ & 
%   \text{subject to } s_{\tau+1} = \widehat{f}_\tau(s_\tau,a_\tau), \ \tau=t,\ldots,t+w-1
% \end{align*}
% where the $\argmax$ operator outputs the optimized action $a_t$ for the $t$-th time step. 

% \textit{Value-based policy.}
% Fix a policy $\pi$ and a value function $\widehat{V}^{\pi}$, the value-based RL policy
% $\pi_{\text{RL}}(s):\mathcal{S}\rightarrow\mathcal{A}$ is defined as
% \begin{align*}
%    \pi_{\text{RL}}(s_t)=& \argmax_{a\in\mathcal{A}}  \widehat{Q}^{\pi}\left(s_t,a\right)\\
%    =& \argmax_{a\in\mathcal{A}} \left(c_t\left(s_t,a\right) + \widehat{V}^{\pi}(f_h(s_t,a))\right)
% \end{align*}

We present theoretical guarantees of the \ouralg policy. 
Consider the error metric $\varepsilon$ defined in~\eqref{eq:def_epsilon}. 
The goal of the OOD-aware scheduling is to achieve (1). near-optimal total cost when the Q-value error $\varepsilon$ is small, and (2). close to the scheduling baseline when the Q-value error $\varepsilon$ is high due to distribution shifts in a stochastic setting. This motivates us to consider the following performance benchmark that generalizes the concept of competitive ratio to evaluate the optimality of scheduling policies. 

% \begin{enumerate}
%     \item \textit{Consistency} Trust the deep policy when the Q-value error $\varepsilon$ is small and achieve near-optimal scheduling cost.  
%     \item \textit{Robustness} Refer to the scheduling baseline when the Q-value error $\varepsilon$ is high due to distribution shifts.
% \end{enumerate}

\begin{definition}
 \label{def:roe}
Given $\mathsf{MDP}(\mathcal{S},\mathcal{A},\nt,\mathbb{P},c)$, the \textbf{ratio of expectations} 
of a policy $\pi=\left(\pi_t:t\in [\nt]\right)$ is defined as $\mathsf{RoE}(\pi)\coloneqq {J(\pi)}/{J^{\star}}$ where $J(\pi)$ is defined in~\eqref{eq:reward} and $J^{\star}$ is the optimal expected cost.
 \end{definition}

Now, we are ready to introduce our definition of consistency and robustness with respect to the ratio of expectations, similar to the growing literature on learning-augmented algorithms~\cite{purohit2018improving,banerjee2020improving,christianson2022chasing}. We write the ratio of expectations $\mathsf{RoE}(\varepsilon)$ of a policy $\pi$ as a function of the {Q}-value advice error $\varepsilon$ in terms of the $\ell_\infty$ norm, defined in~\eqref{eq:def_epsilon}.
\begin{definition}[Consistency and Robustness]
\label{def:tradeoff}
An algorithm $\pi$ is said to be \textbf{$k$-consistent} if its worst-case (with respect to the MDP) ratio of expectations satisfies $\mathsf{RoE}(\varepsilon)\leq k$ for $\varepsilon=0$. On the other hand, it is \textbf{$l$-robust} if $\mathsf{RoE}(\varepsilon)\leq l$ for any $\varepsilon>0$.
\end{definition}

The result below indicates that for the non-stationary dynamics in~\eqref{eq:charging_dynamics}, \ouralg achieves a near-optimal consistency and robustness trade-off. 
{ 
\begin{theorem}
\label{thm:dynamic_regret}
Consider the EV charging problem in Section~\ref{sec:model} with Assumption~\ref{assump:bounded-costs-and-dynamics} and~\ref{assump:uniform-stability}. Suppose the neural policy satisfies Assumption~\ref{ass:Q-Lip}. With the choice of the awareness radius in~\eqref{eq:budget} there exists $\beta>0$ such that \ouralg is $\left(1+\mathcal{O}(\overline{W})\right)$-consistent and $\left( \mathsf{RoE}(\MPC)+\mathcal{O}(\overline{W})+o(1)\right)$-robust where 
\begin{align}
\label{eq:roe_mpc}
    \mathsf{RoE}(\MPC)\leq \frac{2\xi C^2(1 + C^2)(1 + \overline{A}^2 + \overline{B}^2)}{\mu (1 - \overline{\decayfactor})^2}
\end{align}
bounds the ratio of expectations of the MPC baseline (Procedure~\ref{alg:mpc_update}) with $\overline{\decayfactor} \coloneqq\left(({\overline{\sigma} - \underline{\sigma}})/({\overline{\sigma} + \underline{\sigma}})\right)^{\frac{1}{2}}, C \coloneqq \frac{4(\xi + 1 + \overline{A} + \overline{B})}{\underline{\sigma}^2 \cdot \decayfactor},$ and
\begin{align*}
\underline{\sigma} &\coloneqq \min\{\mu, 1\} (\overline{A} + \overline{B} + 1)  \left({\xi}/({2\mu \xi + \mu \sigma^2})\right)^{\frac{1}{2}},\\
\overline{\sigma} &\coloneqq \sqrt{2} (\xi + \overline{A} + \overline{B} + 1).
\end{align*}
\end{theorem}}

% $C$ and $\overline{\lambda}$ are constants defined in Theorem~\ref{thm:dynamic_regret}

% where $\frac{2\xi C^2(1 + C^2)(1 + \overline{A}^2 + \overline{B}^2)}{\mu (1 - \overline{\decayfactor})^2}$
% and $\overline{C}$ is a constant defined as 
% \begin{align}
% \label{eq:constant_C}
%     \overline{C}\coloneqq C(1+C)(\overline{A}+\overline{B})\frac{1}{1-\overline{\lambda}},
% \end{align}

The proof of Theorem~\ref{thm:dynamic_regret} is outlined in Appendix~\ref{app:proof}.
This result highlights a trade-off between the consistency and robustness for learning-augmented algorithms~\cite{purohit2018improving,banerjee2020improving,christianson2022chasing,golowich2022can}. The term $\mathcal{O}(\overline{W})$ with $\overline{W}$ bounding the perturbations $\norm{\ell'_t-\Delta h'_t} \leq \overline{W}$ for all $t\in [\nt]$ in Assumption~\ref{assump:bounded-costs-and-dynamics} is natural and measures the state estimation error in~\eqref{eq:charging_state_estimate}. {Theorem~\ref{thm:dynamic_regret} is technically novel in the following two aspects. First, different from the considered linear time-varying dynamics in~\cite{li2023beyond}, Theorem~\ref{thm:dynamic_regret} deals with nonlinear dynamics specifically designed for adaptive EV charging. Moreover, another nontrivial feature of our model is that the system state $s_t=(e_t\| b_t)\in \mathbb{R}^{\nn}$ (defined in Section~\ref{eq:LQ-ACN}) at each time cannot be directly measured and therefore, it needs to be estimated. Second, in contrast to~\cite{li2023beyond}, we show that the MPC optimization in~\eqref{eq:mpc} for EV charging with nonlinear constraints and state estimation (noting that it is also a practical scheduling algorithm implemented in real charging stations. See~\cite{lee2021adaptive,lee2021acn} for more details) is robust in terms of the Wasserstein distance, leading to a generalized performance guarantee as another main contribution in our algorithm design.
}

% The proofs of Theorem \ref{thm:dynamic_regret}  can be found in Appendix~\ref{app:proof} and Corollary~\ref{coro:algorithm_performance} is a direct implication of Theorem \ref{thm:dynamic_regret} following the proof of Theorem 5.4 in \cite{li2023beyond}.

\section{Case Studies}
\label{sec:exp}

In our case studies, we explore a scenario within the framework of the EV charging model discussed in Section~\ref{eq:LQ-ACN}. In light of the distribution shift of charging patterns observed in Figure~\ref{fig:compare_together}, the charging sessions $\left(\alpha_j,\delta_j,\kappa_j, i\right)$ are formed by realistic user behaviors in the ACN-Data~\cite{lee2019acn}.  To emulate real-world constraints, a charging session is omitted when all charging stations are simultaneously occupied. 

\subsection{Charging Parameters}
We implement the charging dynamics defined in~\eqref{eq:original_dynamics} (see Section~\ref{sec:model}), with system matrices
\begin{align*}
A =&\begin{pmatrix}
 \bigone_{\nm}
  & \rvline & -\Delta\mu\bigone_{\nm} \\
\hline
  \bigzero & \rvline &
 \bigone_{\nm}
\end{pmatrix}, \ B=\begin{pmatrix}
 0 \\
\hline
 \beta\bigone_{\nm}
\end{pmatrix}, \
Q=\bigone_{\nn}, \ R = \alpha \bigone_{\nm},
\end{align*}
where the time span is $10$ mins, (thus $\Delta=1/6$), charging efficiency $\mu=0.8$, control efficiency $\beta=0.2$, $\alpha=0.1$, $\nt=24\times 6$, $\nm=2$, $\nn=4$. 
With the settings above, we encapsulate an OpenAI Gym environment~\cite{brockman2016openai} with action space and state space defined as hyper-rectangles in $\mathbb{R}^2$ and $\mathbb{R}^4$. The control actions $a_{ti}\in [-2,2]$ adjusts the charging rate for all $i\in \{1,\ldots, \nm\}$ and $t\in [\nt]$. The charging state $s_{ti}\in [-100,100]$ for $i\in \{1,\ldots, \nm\}$ and $s_{ti}\in [-6.6,6.6]$ for $i\in \{\nm+1,\ldots,\nn\}$. The charging rates are set the same as the real charging station where ACN-Data~\cite{lee2018large} was collected.

\subsection{Learning Algorithm Settings}

\subsubsection{MPC Baseline $\pi_{\MPC}$}

With the estimated system dynamics $g_{\textsf{MPC}}$ in~\eqref{eq:charging_state_estimate} from the user-input departure time and battery capacity information, we use the MPC policy described in~\eqref{eq:mpc} as the baseline policy in our experiments.

\subsubsection{Machine-Learned Policy $\pi_{\NN}$} We use the Deep Deterministic Policy Gradient (DDPG) algorithm~\cite{silver2014deterministic} to generate machine learned advice and Q-value functions, with hyper-parameters set as (see Table~\ref{table:ddpg} in Appendix~\ref{app:table}).

\subsection{Experiments}

We examine the \ouralg policy by focusing on a scenario where a distribution shift in the charging dynamics~\eqref{eq:original_dynamics} in Section~\ref{sec:model} occurs over the course of training $1,200$ episodes.

\subsubsection{Impact of COVID-19} 
\begin{figure}[ht]
\centering
\includegraphics[width=0.8\linewidth]{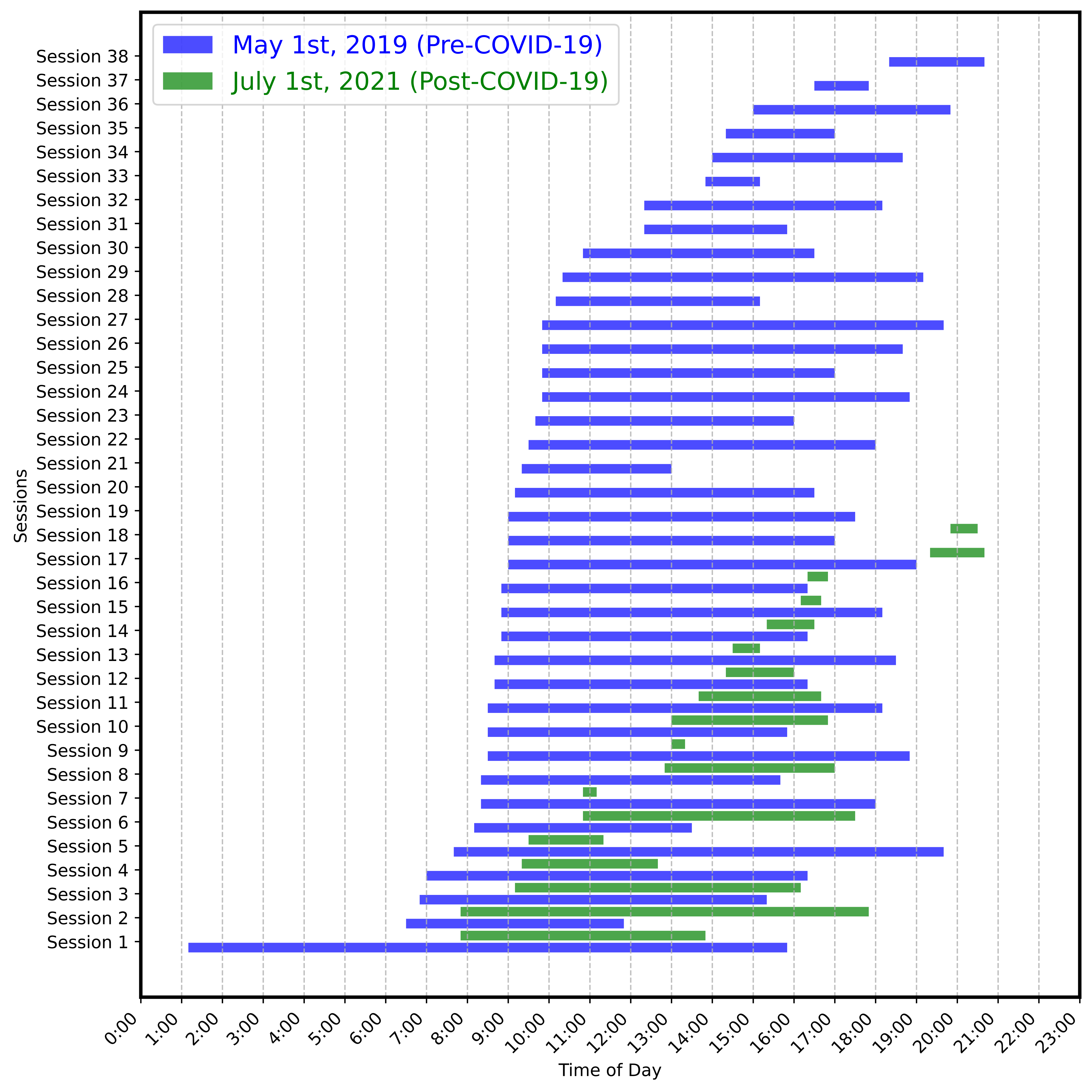}
\caption{Comparison of EV charging sessions on May 1, 2019 (pre-COVID), and July 1, 2021 (post-COVID), illustrating the distribution shift central to this case study. Post-COVID arrival times exhibit a flatter slope, aligning with the reduced density observed in the top-left section of Figure~\ref{fig:compare_together}.}
\label{fig:sessions}
\end{figure}

The distribution shift in charging dynamics can be attributed to two principal events. Firstly, the onset of COVID-19 significantly altered the charging patterns of EV drivers, as evidenced by the data presented in Figure~\ref{fig:data_shift} and Figure~\ref{fig:compare_together}. The neural network for DDPG was initially trained using an OpenAI Gym~\cite{brockman2016openai} environment  developed from ACN-Data, specifically gathered on May 1, 2019, covering $800$ episodes. This training was later supplemented with out-of-distribution data, which was collected on July 1, 2021, to reflect the changing conditions. The contrast in terms of charging session profiles between the data from these two timeframes is shown in Figure~\ref{fig:sessions}.

\subsubsection{Impact of Solar Variation}
% Second, we consider two different solar generation cases.
% For the injection $h_t$  at time $t\in [\nt]$ in~\eqref{eq:original_dynamics}, we treat it as a Gaussian vector so that $h_t = h\times I$ where $h$ is an independent Gaussian random variable.  For the first $800$ episodes, each $h$ is sampled from a normal distribution $\mathcal{N}(\mu,\sigma)$ where $\mu=10$ and $\sigma=0.05$. This represents a case of a sunny day. However, in the last $400$ episodes, we adjust $\mu$ to $0$, representing a rainny day. We normalize the rewards so the MPC policy achieves nearly stable rewards for a better presentation. 
The second aspect of the distribution shift pertains to variations in solar generation. We address solar injection at time \( t \) within the training episodes by modeling it as a Gaussian random vector, such that each injection \( h_t \) is equal to a scalar \( h \) times the identity matrix \( I \), where \( h \) is drawn from an independent Gaussian distribution. For the initial $800$ episodes, \( h \) is sampled from a normal distribution \( \mathcal{N}(\mu,\sigma^2) \) with mean \( \mu=10 \) and standard deviation \( \sigma=0.05 \), corresponding to a sunny day. Conversely, for the remaining $400$ episodes, we set \( \mu \) to $0$, representing a rainy day. Reward normalization was conducted to ensure that the MPC policy yielded consistently stable rewards across these variations for more effective analysis.

\subsubsection{Experimental Results}

% For this case study, Figure~\ref{fig:reward} depicts the reward trajectory following a distributional shift indicative of changes from a pre-COVID/sunny period to a post-COVID/rainy one, across different $\beta$ values.

\begin{figure}[ht]
\centering
\includegraphics[width=0.45\textwidth]{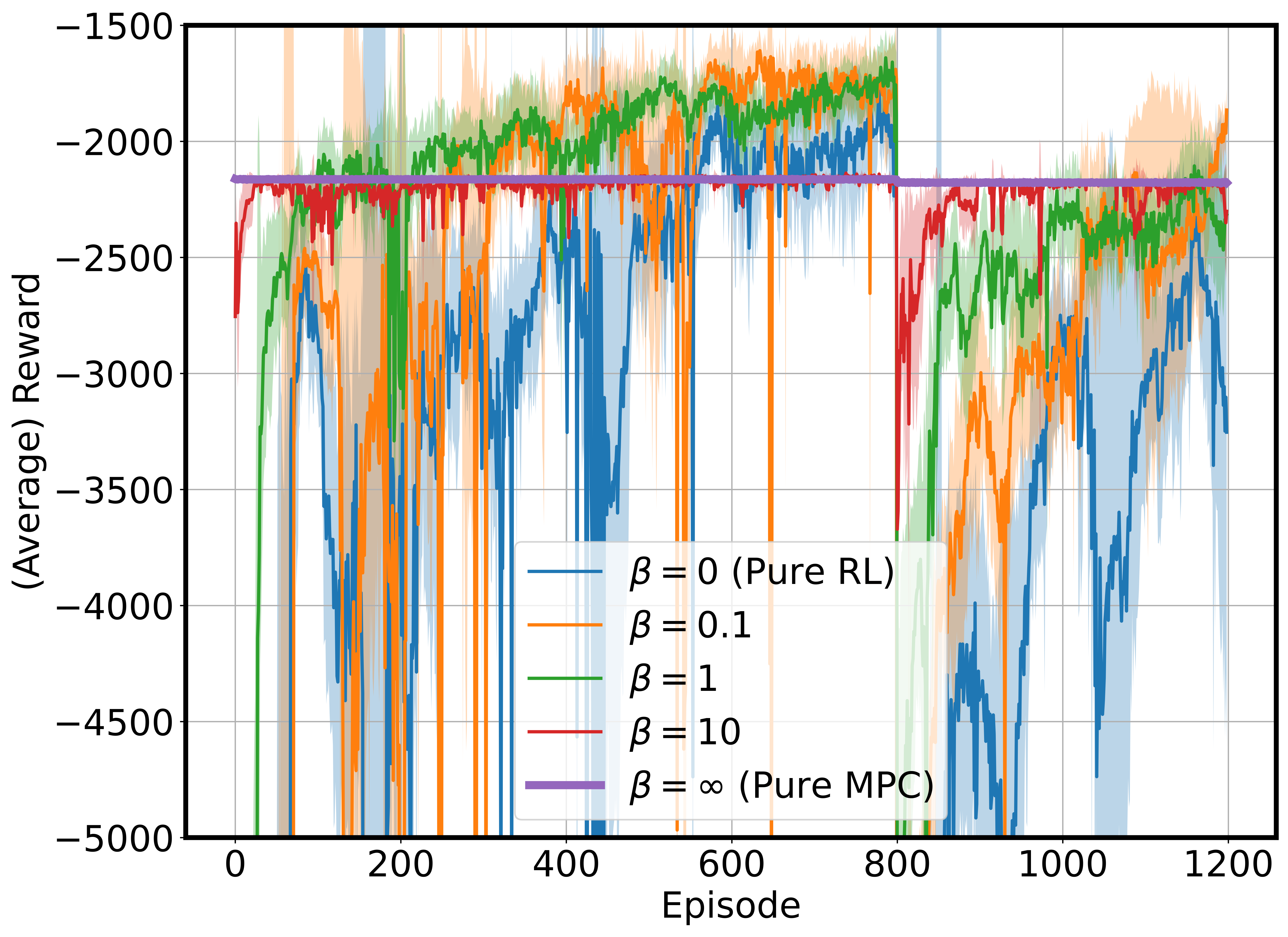}
\caption{We evaluate the resilience of awards against distribution shifts by varying the tuning parameter $\beta$. The plot showcases average awards for $\beta$ values of $0, 0.1, 1, 10$, and $\infty$—the latter representing the direct application of the MPC baseline—within \ouralg's robustness framework. The shaded regions indicate the standard deviation across $10$ independent experiments. { The implemented MPC (Pure MPC) is identical to the scheduling algorithms currently used at real EV charging stations at Caltech and JPL garages. Importantly, this implementation does not rely on historical charging data, which implies that its reward remains constant with respect to the training episode. }}
\label{fig:reward}
\end{figure}

In this case study, Figure~\ref{fig:reward} illustrates the trajectory of rewards after a distributional shift, representative of the transition from a pre-COVID/sunny period to a post-COVID/rainy period, across various values of the hyper-parameter $\beta$. {Numerical results of the average rewards and standard deviation after convergence for both pre-COVID and post-COVID periods are summarized in Table~\ref{table:reward}.}

\begin{figure}[ht]
\centering
\includegraphics[width=0.45\textwidth]{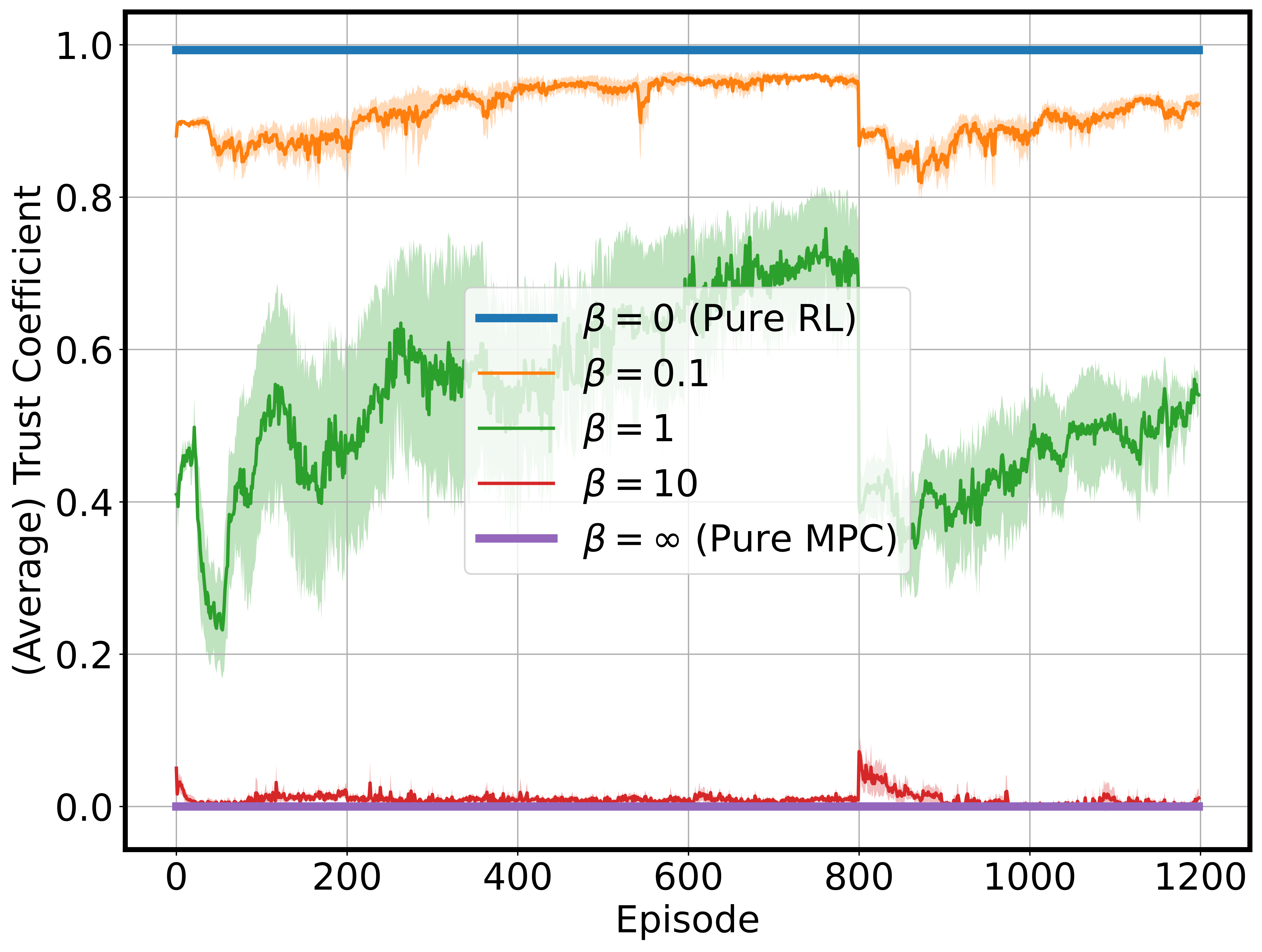}
\caption{Impact of hyper-parameter \(\beta\) on projection radii $R_t$ over time intervals \(t \in [\nt]\). The shaded area illustrates the standard deviation from $10$ independent trials. The graph demonstrates a consistent decrease in the average trust coefficient \(\lambda(R_t)\) with increasing values of \(\beta\), indicating a negative correlation between \(\beta\) and the reliability of the projection radii in \ouralg for EV charging scheduling.}
\label{fig:lambda}
\end{figure}

\begin{figure}[ht]
\centering
\includegraphics[width=0.45\textwidth]{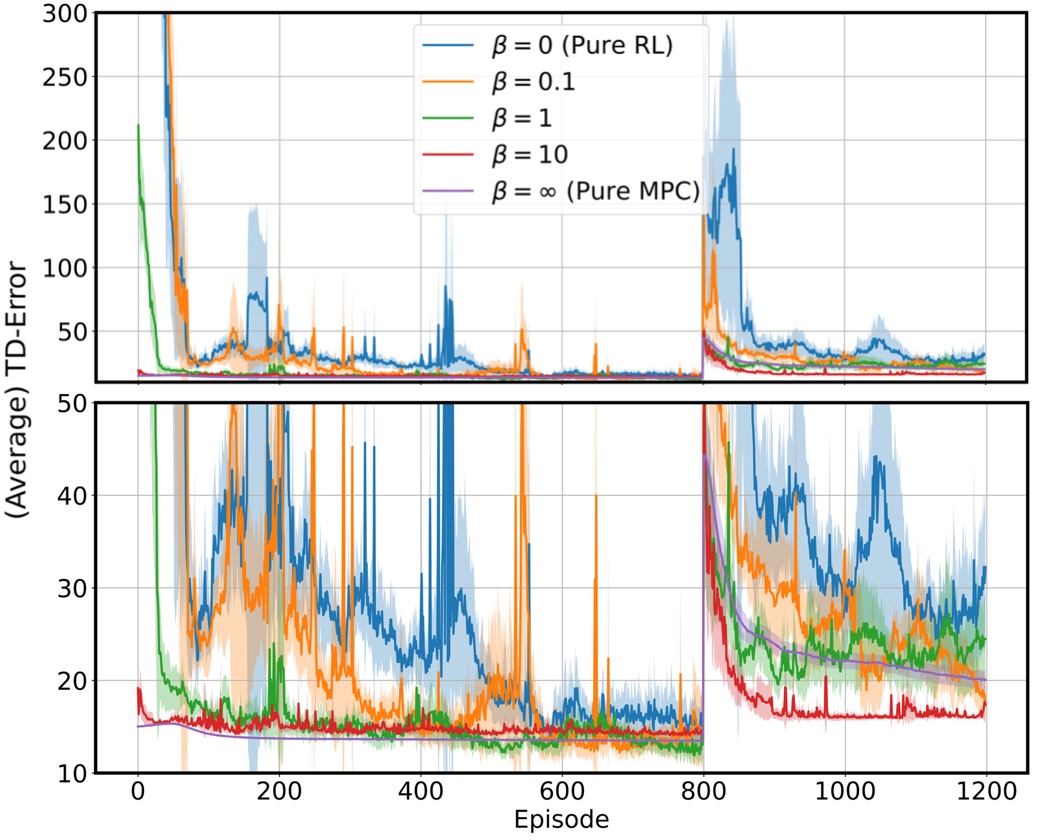}
\caption{Average approximate TD-error (see~\eqref{eq:td_error_approximate}) as a function of the hyper-parameter \(\beta\) in the awareness radius (see~\eqref{eq:budget})of \ouralg for EV charging scheduling. The shaded area represents the range of one standard deviation computed from $10$ independent tests.}
\label{fig:td}
\end{figure}

Given the MPC policy $\pi_{\MPC}\coloneqq \{\overline{\pi}_t : t \in [\nt]\}$ and the machine-learned neural network (NN) policy $\pi_{\NN}\coloneqq \{\widetilde{\pi}_t : t \in [\nt]\}$, the action selected by \ouralg at each time step $t \in [\nt]$ is given by:
$
    u_t = \lambda(R_t)\widetilde{\pi}_t(x_t) + (1 - \lambda(R_t))\overline{\pi}_t(x_t),
$
where $\lambda(R_t) \coloneqq \min\left\{1, {R_t}/{\|\widetilde{\pi}_t(x_t) - \overline{\pi}_t(x_t)\|_2}\right\}$ is defined as the \textit{trust coefficient}, ranging between $0$ and $1$. When the trust coefficient is higher, greater confidence is placed in the neural network (NN) policy, resulting in the use of a larger radius $R_t$ in the projection step of \ouralg. Here, $R_t$ represents an \textit{awareness radius}, as defined in Equation~\eqref{eq:budget}. Figure~\ref{fig:lambda} illustrates the behavior of $\lambda(R_t)$, averaged over all time steps and tests. Additionally, Figure~\ref{fig:td} displays the trend of the approximate TD-error for different $\beta$ values, also averaged over all time steps and tests. Notably, when $\beta = 1$, the approximate TD-error exhibits a pronounced convergence, coinciding with high average rewards as shown in Figure~\ref{fig:reward}.

\begin{table}
\footnotesize
  \centering
  \renewcommand{\arraystretch}{1.3}
  \begin{tabular}{p{0.8cm}|c|c|c|c}
\specialrule{.11em}{.1em}{.1em} 
    \multirow{2}{4cm}{$\beta$} & \multicolumn{2}{c|}{\textbf{Average Rewards} ($k$)} & \multicolumn{2}{c}{\textbf{Average SD}}\\
    % \hline
    % \textbf{Inactive Modes} & \textbf{Description}\\
    \cline{2-5}
    & \textit{pre-COVID}  & \textit{post-COVID} & \textit{pre-COVID}  & \textit{post-COVID}\\
    %\hhline{~--}
    \hline
    $0$ & $-2.380$ & $-3.780$ & $906.856$ & $1770.791$ \\ \hline
    $0.1$ & $-1.813$ & $-2.386$ & $447.444$ & $878.278$ \\ \hline
    $1$ & $-1.827$ & $-2.320$ & $294.858$ & $418.767$   \\ \hline
    $10$ & $-2.155$ & $-2.192$ & $41.532$ & $70.328$   \\ \hline
    $\infty$ & $-2.153$ & $-2.153$ & $-$ & $-$   \\ 
    \specialrule{.11em}{.1em}{.1em} 
  \end{tabular}
  \caption{{Average rewards and standard deviation (SD) for pre-COVID (episodes $600$ to $800$) and post-COVID (episodes $1,000$ to $1,200$) periods with $\beta$ values of $0,0.1,1,10$ and $\infty$.}}
  \label{table:reward}
\end{table}

% With $\beta$ set to $10$, the results altogether suggest a near-optimal balance between consistency and robustness. It is worth highlighting that \ouralg maintains high average rewards prior to the shift and demonstrates rapid adaptation post-shift, underscoring the algorithm's resilience and effectiveness. This performance is achieved without specific fine-tuning of $\beta$, mirroring the initial experimental results.

With \(\beta\) set to \(1\), the results collectively indicate a near-optimal trade-off between consistency and robustness. Remarkably, \ouralg reduces the variability of the pure RL algorithm and maintains high average rewards prior to the distribution shift and exhibits prompt adaptation thereafter, verifying the algorithm's resilience and efficacy. It is worth highlighting that we did not actively fine-tune \(\beta\), underscoring the robustness and adaptability of our approach. 
Moreover, we compare against MPC as a representative baseline of classic charging algorithms, since as thoroughly evaluated in the ACN infrastructure~\cite{lee2021acn,lee2021adaptive}, MPC-type scheduling has the best performance among baseline methods including Round Robin, First Come First Serve (FCFS), Earliest Deadline First (EDF), Least Laxity First (LLF), closely approximating the offline optimal results.

\section{Concluding Remarks}
\label{sec:conclusion}
In this paper, we present an OOD-aware policy, \ouralg, to tackle the challenge of learning to charge EVs under OOD conditions.
Experimental results show that \ouralg significantly improves the effectiveness and reliability of EV charging scheduling in unpredictable real-world environments, particularly in OOD contexts. It has several limitations
that motivate extensions. 
In future work, it would be interesting to assess the performance of the proposed approach under more stringent real-world constraints such as three-phase network models and customized user preferences. Additionally, we plan to implement the proposed OOD-aware algorithm at real charging sites with solar cells installed to verify the applicability of the proposed algorithm.
% Additionally, we aim to test the scalability of OOD-aware policy on larger-scale EV charging networks and assess its computational efficiency in practice.

\appendix

\subsection{Hyperparameters in Experiments}
\label{app:table}

We present the detailed hyperparameters in Table~\ref{table:ddpg}.

\begin{table}[ht]
\footnotesize
\renewcommand{\arraystretch}{1.3}
    \centering
       \caption{Hyper-parameters used in the DDPG-based procedure \textsc{NN-Update} (Procedure~\ref{alg:nn_update}).}
    \begin{tabular}{l|l}
\specialrule{.11em}{.1em}{.1em} 
   \textbf{Hyper-Parameter}  & \textbf{Value}  \\
   \hline
   Maximal number of episodes & $10^3$ \\
   Episode length $\nt$ & $144$ \\
    % Discount factor & $1.0$\\
   Network learning rate $\Gamma$ & $10^{-3}$ \\
   % Soft target update parameter & $10^{-3}$\\
   Replay buffer size $|\mathcal{D}|$ & $10^6$ \\
   Mini-batch size $|B|$ &  $128$\\
\specialrule{.11em}{.1em}{.1em} 
    \end{tabular}
  \label{table:ddpg}
\end{table}

\subsection{Proof of Theorem~\ref{thm:dynamic_regret}}
\label{app:proof}

\textit{Step 1. Wasserstein Robustness.}
Applying Theorem C.1 in~\cite{li2023beyond} that holds for a general MDP model, it suffices to verify that the MPC scheduling baseline $\pi_{\MPC}$ satisfies the following Wasserstein robustness condition.

\begin{definition}[$r$-locally $p$-Wasserstein robustness~\cite{li2023beyond}] \label{def:robust}  A policy $\overline{\pi}=(\pi_t:t\in [\nt])$ is \textbf{$r$-locally $p$-Wasserstein-robust} if for any $0\leq t_1\leq t_2<\nt$ and 
 state-action distributions $\rho,\rho'$ such that $W_p(\rho,\rho')\leq r$,
% whose support is $\bigcup_{y\in \mathcal{X}(\overline{\pi})}\mathcal{B}(y,r)$ 
for some radius $r>0$, \begin{align} 
\label{eq:def_robustness}
{W}_p\left(\rho_{t_1:t_2}(\rho),\rho_{t_1:t_2}(\rho')\right)\leq & h(t_2-t_1) {W}_p\left(\rho,\rho'\right)\end{align} for  some function $h:[\nt]\rightarrow \mathbb{R}_+$ such that $\sum_{t\in [\nt]}h(t)\leq \overline{C}$ for some constants $\overline{C}>0$.  \end{definition}

We first argue that the verification of Wasserstein robustness of the MPC scheduling baseline $\pi_{\MPC}$ can be {generalized} for a reparameterized dynamics in~\eqref{eq:charging_dynamics} of the charging dynamics~\eqref{eq:original_dynamics} in Section~\ref{sec:model}, with state and action-dependent perturbations $(w_t(s_t,a_t): t\in [\nt])$. To see this, let $\Delta$ be the time interval between state updates. Denote by
\begin{subequations}
\begin{align}
\label{eq:arrival_set}
    \mathcal{A}_i &\coloneqq\left\{(j,t):\alpha_j\leq t\leq \alpha_j +\Delta, c_j^4=i\right\}, \ \text{and}\\
\label{eq:departure_set}
    \mathcal{D}_i &\coloneqq\left\{(j,t):\delta_j\leq t\leq \delta_j +\Delta, c_j^4=i\right\},
\end{align}
\end{subequations}
the sets of session labels with the corresponding time indices when the sessions start and end at charger $i$ respectively. Below we introduce a concrete MPC formulation for EV charging, based on the following charging dynamics as a special case of $\mathsf{MDP}(\mathcal{S},\mathcal{A},\nt,\mathbb{P},c)$:
\begin{align}
\label{eq:charging_dynamics}
     s_{t+1}=& \underbrace{A_ts_t+B_ta_t}_{\textit{Battery Dynamics}} + \underbrace{w_t(s_t,a_t)}_{\textit{Perturbations}},
\end{align}
where $(w_t:t\in [\nt])$ can be stochastic perturbations caused by human charging behavior, explained below.
The matrices $(A_t:t\in [\nt])$ and $(B_t:t\in [\nt])$ are known battery dynamics used to model the time-varying behaviors of a non-ideal battery~\cite{lee2021adaptive}. Based on above, provided with all session information, with the sets of session labels defined in~\eqref{eq:arrival_set} and~\eqref{eq:departure_set}, an example of the perturbations can be expressed as, for any $i=1,\ldots,\nm$,
\begin{align*}
w_{ti}(s_t,a_t) \coloneqq & \mathbb{I}\left((j,t)\in\mathcal{A}_i\right) \kappa_j -\mathbb{I}\left((j,t)\in\mathcal{D}_i\right)\left(A_{ti}s_t+b_{ti}a_t\right) \\
-& \mathbb{I}\left(A_{ti} s_t+b_{ti} a_t< -\overline{c}\right)\left(A_{ti} s_t+b_{ti} a_t +\overline{c}\right)\\
- & \mathbb{I}\left(A_{ti} s_t+b_{ti} a_t >  \overline{c}\right)\left(A_{ti} s_t+b_{ti} a_t -\overline{c}\right), 
\end{align*}
and for any $i=\nm+1,\ldots,\nn$,
\begin{align*}
w_{ti}(s_t,a_t) \coloneqq &  g_{\mathcal{A}}(a_t)_i - B_{ti} a_t + \Delta h_{ti} \\
+ &\mathbb{I}\left(s_{ti}>\overline{b}\right)\left(\overline{b}-s_{ti}\right) -\mathbb{I}\left(s_{ti}<0\right)s_{ti}   \\
+ & \mathbb{I}\left(\sum_{i=\nm+1}^{\nn}s_{ti}>\gamma \right) \left(\frac{\gamma}{\sum_{i=\nm+1}^{\nn}s_{ti}} -1\right)s_{ti},
\end{align*}
% \begin{align}
% \nonumber
% w_t^{i}(s_t,a_t)\coloneqq 
% \begin{cases}
% \kappa_j
% & \text{if } (j,t)\in\mathcal{A}_i \\
% -A_t^{i}s_t-b_{ti}a_t
% & \text{if } (j,t)\in\mathcal{D}_i \\
% -A_t^{i}s_t-b_{ti}a_t
% & \text{if } A_t^{i} s_t+b_{ti} a_t<0 \\
% b_{ti} \left(\overline{a} -a_t\right)
% & \text{if } a^{i}_t>\overline{a} \\
% b_{ti} \left(\frac{\gamma}{\sum_{i=1}^{\nn}a_t^{i}} -1\right)a_t
% & \text{if } \sum_{i=1}^{\nn}a_t^{i}>\gamma \\
% 0
% & \text{otherwise }
%  \end{cases}
%  \end{align}
where $A_{ti}$ and $B_{ti}$ are the $i$-th rows of matrices $A$ and $B$, respectively. 
Note that in~\eqref{eq:charging_dynamics},  {the events that trigger a state-dependent perturbation are} (1). the installed and EV batteries are full; (2).  the charging rates reach at either the line or individual charging limit; and (3). the EV departs. { Consider two distinct states $s_t$ and $s_t'$.} For the first and second events, this will only make the two SoC states $e_{t'}$ and $e_{t'}'$ (charging rate states $b_{t'}$ and $b_{t'}'$) at time $t'$ corresponding to initial states $e_{t}$ and $e_{t}'$ ($b_{t}$ and $b_{t}'$) closer to each other since the SoC states and charging rate states are constrained in some bounded spaces. Finally, for the third event, {the corresponding coordinates $e'_{t'i}$} and $e_{t'i}$ with respect to the charger $i$ will be the same since they will be initialized to zeros after the EVs depart. The remaining proof is similar to that of Theorem A.1. in~\cite{li2023beyond}. 

Moreover, another major challenge we address here is that in our setting generalizes the predictions of perturbations $(\widetilde{w}_t:t\in [\nt])$ to possibly non-zero user-inputs while in~\cite{li2023beyond}, the predictions are set to all zeros. Define an operator $\psi_{t:t'}:\mathcal{S}^{t'-t+1}\times \mathbb{R}^{\nn\times\nn}\rightarrow \mathcal{A}^{t'-t}$ as the optimal actions of the following optimization. To simplify the notation, we use the shorthand $\psi_{t, t'}(s_t; P) \coloneqq \psi_{t, t'}(s_t, (\widetilde{w}_{\tau|t}:\tau\in [t:t'-1]); P)$, and define
\[\psi_{t, t'}(s_t) \coloneqq \begin{cases}
\psi_{t, t'}(s_t; P_{t'}) &\text{ if } t' < \nt-1,\\
\psi_{t, t'}(s_t; Q_{t'}) &\text{ if } t' = \nt-1.
\end{cases}\]

By the KKT conditions, we see that for any $t \in [\nt]$, the predictive optimal solution $\psi_{t, \nt-1}(s_t)$ is given by
\begin{align}\nonumber
    \left(\begin{array}{c}
         s_{t|t}\\
         a_{t|t}\\
         \vdots\\
         s_{\nt-1|t}\\
         \hline
         \eta_{t|t}\\
         \vdots\\
         \eta_{\nt-1|t}
    \end{array}\right) = 
    \left(\begin{array}{c|c}
        \Gamma_{t, \nt-1} & (\Phi_{t, \nt-1})^\top \\
        \hline
        \Phi_{t, \nt-1} & \mathbf{0}
    \end{array}\right)^{-1} \left(\begin{array}{c}
         0\\
         \vdots\\
         0\\
         \hline
         s_t\\
         \widetilde{w}_{t|t}\\
         \vdots\\
         \widetilde{w}_{\nt-1|t}
    \end{array}\right).
\end{align}
Therefore, using the relationship above and applying Lemma 5 and 6 in~\cite{li2023beyond}, we show that the MPC baseline is Wasserstein robust with a constant $\overline{C}=C(1+C)(\overline{A}+\overline{B})\frac{1}{1-\overline{\lambda}}$. Moreover, the baseline has a bounded competitive ratio $$\frac{2\xi C^2(1 + C^2)(1 + \overline{A}^2 + \overline{B}^2)}{\mu (1 - \overline{\decayfactor})^2}$$ for the worst case perturbations.

\vspace{2pt}

\textit{Step 2. Consistency Bound.} Now we show that the per-step expected dynamic regret of \ouralg is less than $\mathbb{E}\left[\mu_t\right] +L_Q\mathbb{E}\left[\eta_t\left(\widetilde{s}_t\right)+2\overline{W}-R_t\right]$, {i.e.,
\begin{align}
\label{eq:app_consistency}
    J(\pi) - J^{\star}\leq \mathbb{E}\left[\mu_t\right] +L_Q\mathbb{E}\left[\eta_t\left(\widetilde{s}_t\right)+2\overline{W}-R_t\right],
\end{align}
where $J^{\star}$ denotes the optimal expected cost defined in Definition~\ref{def:roe}.}
We apply the projection lemma (Lemma 3) in \cite{li2023beyond}, and note that the learned Q-value function and the induced trajectory of states and actions satisfy
\begin{align}
\widetilde{Q}_{t}\left(s_t,a_t\right) 
\nonumber
\leq & \widetilde{Q}_{t}\left(s_t,\widetilde{\pi}_t\left(\widetilde{s}_t\right)\right)+  L_Q \left\|\widetilde{\pi}_t\left(\widetilde{s}_t\right)-a_t\right\|\\
% \nonumber
%    \leq & \widetilde{Q}_{t}\left(\widetilde{s}_t,\widetilde{\pi}_t\left(\widetilde{s}_t\right)\right)+  L_Q \left\|\widetilde{\pi}_t\left(\widetilde{s}_t\right)-a_t\right\| + 2L_Q \overline{W} \\
% \nonumber
%       \leq  &  \widetilde{Q}_{t}\left(\widetilde{s}_t,\widetilde{\pi}_t\left(\widetilde{s}_t\right)\right) +  L_Q\left(1-\lambda\left(R_t\right)\right)\eta_t\left(\widetilde{s}_t\right) + 2L_Q \overline{W} \\
     %  \nonumber
     % = &  \widetilde{Q}_{t}\left(\widetilde{s}_t,\widetilde{\pi}_t\left(\widetilde{s}_t\right)\right) + L_Q\left(1-\min\left\{1,\frac{R_t}{\eta_t\left(\widetilde{s}_t\right)}\right\}\right)\eta_t\left(\widetilde{s}_t\right)+ 2L_Q \overline{W} \\
\nonumber
   \leq    & \widetilde{Q}_{t}\left(s_t,\widetilde{\pi}_t\left(\widetilde{s}_t\right)\right) + L_Q\left(\eta_t\left(s_t\right)-R_t\right) + 2L_Q \overline{W} 
\end{align}
where $\widetilde{s}_t$ is the estimated state at each time $t$, which satisfies that $\|s_t-\widetilde{s}_t\|\leq 2\overline{W}$ and $\widetilde{a}_t$ is the action obtained by minimizing the Q-value function $\widetilde{Q}_t$.
Therefore, the result can be derived by using
\begin{align}
 \nonumber
Q_{t}^{\star}\left(s_t,a_t\right)-\inf_{a'\in\mathcal{A}} Q_t^{\star}\left(s_t,a'\right)
 \leq &  L_Q\left[\eta_t\left(s_t\right)-R_t\right]^{+}+\mu_t
\end{align}
where $\eta_t\coloneqq \|\overline{a}_t-\widetilde{a}_t\|$, $\mu_t\coloneqq\zeta_{t}^{V}-\zeta_{t}^{Q}$,
$\zeta_{t}^{V} \coloneqq\widetilde{Q}_{t}(s_t,\widetilde{a}_{t})-Q_{t}^{\star}(s_t,a_{t}^{\star})$, and $\zeta_{t}^{Q} \coloneqq\widetilde{Q}_{t}(s_t,a_t)-Q_{t}^{\star}(s_t,a_t)$. 

% Here $Q_{t}^{\star}$ denotes the optimal value function.

\vspace{3pt}

\textit{Step 3. Robustness Bound.}
Since the cost functions $(c_t:t\in [\nt])$  are Lipschitz continuous with a Lipschitz constant $L_{\mathrm{C}}$, using the Kantorovich Rubinstein duality theorem, since $ {W}_p(\mu,\nu)\leq  {W}_q(\mu,\nu)$ for all $1\leq p\leq q<\infty$, for all $t\in [\nt]$,
\begin{align}
\nonumber
J(\pi) - J(\overline{\pi}) =&  \sum_{t\in [\nt]} 
\mathbb{E}\left[c_t\left(s,a\right)\right]-\mathbb{E}\left[c_t\left(s,a\right)\right] \\
 \nonumber
% \label{eq:kantorovich}
\leq & \sum_{t\in [\nt]} \sup_{\|f\|_{L}\leq L_{\mathrm{C}}}\mathbb{E}\left[f\left(s,a\right)\right]-\mathbb{E}\left[f\left(s,a\right)\right]\\
\label{eq:robustness_final}
\leq & L_{\mathrm{C}} \overline{C}\sum_{t\in [\nt]} \mathbb{E}\left[R_{t} + 2C\overline{W}\right],
\end{align}
where $\|\cdot\|_L$ denotes the Lipschitz semi-norm and the supremum is over all Lipschitz 
 continuous functions $f$ with a Lipschitz constant $ L_{\mathrm{C}}$. We have applied Lemma 5 in \cite{li2023beyond} such that the MPC control policy can be rewritten as $a_t = \overline{K}_t s_t$ and $\norm{\overline{K}_t} \leq C$ for all $t \in [\nt]$ where $C$ is defined in Theorem~\ref{thm:dynamic_regret}. Since $\|s_t-\widetilde{s}_t\|\leq 2\overline{W}$, we obtain $\overline{\pi}_t(s_t)-\overline{\pi}_t(\widetilde{s}_t)\leq 2C\overline{W}$, which implies~\eqref{eq:robustness_final}. 
 
 { Following the routine arguments in the consistency and robustness analysis for Theorem 5.4 in~\cite{li2023beyond} (see Appendix D.2 therein), the design of the awareness radius $R_t$ in~\eqref{eq:budget} based on the TD-error guarantees that $\mathbb{E}[\eta_t(\widetilde{s}_t)-R_t]=0$ and $\mathbb{E}[R_t]=o(\nt)$. Since $J^{\star}=\Omega(\nt)$,~\eqref{eq:app_consistency} implies that \ouralg is $\left(1+\mathcal{O}(\overline{W})\right)$-consistent (corresponding to the case when $\varepsilon=0$, so $\mu_t=0$ for all $t\in [\nt]$). 
 
 Finally, noting that $\mathsf{RoE}(\MPC)\coloneqq {J(\overline{\pi})}/{J^{\star}}$,  evaluating~\eqref{eq:robustness_final} for all $\varepsilon>0$   implies that \ouralg is $\left( \mathsf{RoE}(\MPC)+\mathcal{O}(\overline{W})+o(1)\right)$-robust where $\mathsf{RoE}(\MPC)$ is bounded from above in~\eqref{eq:roe_mpc}.}

\bibliographystyle{IEEEtran}
\bibliography{main.bib}

\begin{thebibliography}{10}
\providecommand{\url}[1]{#1}
\csname url@samestyle\endcsname
\providecommand{\newblock}{\relax}
\providecommand{\bibinfo}[2]{#2}
\providecommand{\BIBentrySTDinterwordspacing}{\spaceskip=0pt\relax}
\providecommand{\BIBentryALTinterwordstretchfactor}{4}
\providecommand{\BIBentryALTinterwordspacing}{\spaceskip=\fontdimen2\font plus
\BIBentryALTinterwordstretchfactor\fontdimen3\font minus \fontdimen4\font\relax}
\providecommand{\BIBforeignlanguage}[2]{{%
\expandafter\ifx\csname l@#1\endcsname\relax
\typeout{** WARNING: IEEEtran.bst: No hyphenation pattern has been}%
\typeout{** loaded for the language `#1'. Using the pattern for}%
\typeout{** the default language instead.}%
\else
\language=\csname l@#1\endcsname
\fi
#2}}
\providecommand{\BIBdecl}{\relax}
\BIBdecl

\bibitem{chen2017dynamic}
Q.~Chen, F.~Wang, B.-M. Hodge, J.~Zhang, Z.~Li, M.~Shafie-Khah, and J.~P. Catal{\~a}o, ``Dynamic price vector formation model-based automatic demand response strategy for pv-assisted ev charging stations,'' \emph{IEEE Transactions on Smart Grid}, vol.~8, no.~6, pp. 2903--2915, 2017.

\bibitem{yan2018optimized}
Q.~Yan, B.~Zhang, and M.~Kezunovic, ``Optimized operational cost reduction for an ev charging station integrated with battery energy storage and pv generation,'' \emph{IEEE Transactions on Smart Grid}, vol.~10, no.~2, pp. 2096--2106, 2018.

\bibitem{chics2016reinforcement}
A.~Chi{\c{s}}, J.~Lund{\'e}n, and V.~Koivunen, ``Reinforcement learning-based plug-in electric vehicle charging with forecasted price,'' \emph{IEEE Transactions on Vehicular Technology}, vol.~66, no.~5, pp. 3674--3684, 2016.

\bibitem{wan2018model}
Z.~Wan, H.~Li, H.~He, and D.~Prokhorov, ``Model-free real-time ev charging scheduling based on deep reinforcement learning,'' \emph{IEEE Transactions on Smart Grid}, vol.~10, no.~5, pp. 5246--5257, 2018.

\bibitem{li2019constrained}
H.~Li, Z.~Wan, and H.~He, ``Constrained ev charging scheduling based on safe deep reinforcement learning,'' \emph{IEEE Transactions on Smart Grid}, vol.~11, no.~3, pp. 2427--2439, 2019.

\bibitem{li2021learning}
T.~Li, B.~Sun, Y.~Chen, Z.~Ye, S.~H. Low, and A.~Wierman, ``Learning-based predictive control via real-time aggregate flexibility,'' \emph{IEEE Transactions on Smart Grid}, vol.~12, no.~6, pp. 4897--4913, 2021.

\bibitem{fachrizal2022optimal}
R.~Fachrizal, M.~Shepero, M.~{\AA}berg, and J.~Munkhammar, ``Optimal pv-ev sizing at solar powered workplace charging stations with smart charging schemes considering self-consumption and self-sufficiency balance,'' \emph{Applied Energy}, vol. 307, p. 118139, 2022.

\bibitem{li2022ev}
S.~Li, W.~Hu, D.~Cao, Z.~Zhang, Q.~Huang, Z.~Chen, and F.~Blaabjerg, ``Ev charging strategy considering transformer lifetime via evolutionary curriculum learning-based multiagent deep reinforcement learning,'' \emph{IEEE Transactions on Smart Grid}, vol.~13, no.~4, pp. 2774--2787, 2022.

\bibitem{lee2021adaptive}
Z.~J. Lee, G.~Lee, T.~Lee, C.~Jin, R.~Lee, Z.~Low, D.~Chang, C.~Ortega, and S.~H. Low, ``Adaptive charging networks: A framework for smart electric vehicle charging,'' \emph{IEEE Transactions on Smart Grid}, vol.~12, no.~5, pp. 4339--4350, 2021.

\bibitem{hendrycks2016baseline}
D.~Hendrycks and K.~Gimpel, ``A baseline for detecting misclassified and out-of-distribution examples in neural networks,'' in \emph{International Conference on Learning Representations}, 2016.

\bibitem{teney2020value}
D.~Teney, E.~Abbasnejad, K.~Kafle, R.~Shrestha, C.~Kanan, and A.~Van Den~Hengel, ``On the value of out-of-distribution testing: An example of goodhart's law,'' \emph{Advances in neural information processing systems}, vol.~33, pp. 407--417, 2020.

\bibitem{lee2018large}
Z.~J. Lee, D.~Chang, C.~Jin, G.~S. Lee, R.~Lee, T.~Lee, and S.~H. Low, ``Large-scale adaptive electric vehicle charging,'' in \emph{2018 IEEE International Conference on Communications, Control, and Computing Technologies for Smart Grids (SmartGridComm)}.\hskip 1em plus 0.5em minus 0.4em\relax IEEE, 2018, pp. 1--7.

\bibitem{li2021coordinating}
Y.~Li, M.~Han, Z.~Yang, and G.~Li, ``Coordinating flexible demand response and renewable uncertainties for scheduling of community integrated energy systems with an electric vehicle charging station: A bi-level approach,'' \emph{IEEE Transactions on Sustainable Energy}, vol.~12, no.~4, pp. 2321--2331, 2021.

\bibitem{kacperski2022impact}
C.~Kacperski, R.~Ulloa, S.~Klingert, B.~Kirpes, and F.~Kutzner, ``Impact of incentives for greener battery electric vehicle charging--a field experiment,'' \emph{Energy Policy}, vol. 161, p. 112752, 2022.

\bibitem{arias2017robust}
N.~B. Arias, A.~Tabares, J.~F. Franco, M.~Lavorato, and R.~Romero, ``Robust joint expansion planning of electrical distribution systems and ev charging stations,'' \emph{IEEE Transactions on Sustainable Energy}, vol.~9, no.~2, pp. 884--894, 2017.

\bibitem{lee2019acn}
Z.~J. Lee, T.~Li, and S.~H. Low, ``Acn-data: Analysis and applications of an open ev charging dataset,'' in \emph{Proceedings of the Tenth ACM International Conference on Future Energy Systems}.\hskip 1em plus 0.5em minus 0.4em\relax ACM, 2019, pp. 139--149.

\bibitem{lee2021acn}
Z.~J. Lee, S.~Sharma, D.~Johansson, and S.~H. Low, ``Acn-sim: An open-source simulator for data-driven electric vehicle charging research,'' \emph{IEEE Transactions on Smart Grid}, vol.~12, no.~6, 2021.

\bibitem{purohit2018improving}
M.~Purohit, Z.~Svitkina, and R.~Kumar, ``Improving online algorithms via ml predictions,'' \emph{Advances in Neural Information Processing Systems}, vol.~31, 2018.

\bibitem{qian2023impact}
K.~Qian, R.~Fachrizal, J.~Munkhammar, T.~Ebel, and R.~C. Adam, ``The impact of considering state-of-charge-dependent maximum charging powers on the optimal electric vehicle charging scheduling,'' \emph{IEEE Transactions on Transportation Electrification}, vol.~9, no.~3, pp. 4517--4530, 2023.

\bibitem{chen2022reinforcement}
X.~Chen, G.~Qu, Y.~Tang, S.~Low, and N.~Li, ``Reinforcement learning for selective key applications in power systems: Recent advances and future challenges,'' \emph{IEEE Transactions on Smart Grid}, vol.~13, no.~4, pp. 2935--2958, 2022.

\bibitem{su2023electric}
S.~Su, Y.~Li, K.~Yamashita, M.~Xia, N.~Li, and K.~A. Folly, ``Electric vehicle charging guidance strategy considering “traffic network-charging station-driver” modeling: A multi-agent deep reinforcement learning based approach,'' \emph{IEEE Transactions on Transportation Electrification}, 2023.

\bibitem{dong2023multi}
J.~Dong, A.~Yassine, A.~Armitage, and M.~S. Hossain, ``Multi-agent reinforcement learning for intelligent v2g integration in future transportation systems,'' \emph{IEEE Transactions on Intelligent Transportation Systems}, 2023.

\bibitem{yang2024multiagent}
X.~Yang, T.~Cui, H.~Wang, and Y.~Ye, ``Multiagent deep reinforcement learning for electric vehicle fast charging station pricing game in electricity-transportation nexus,'' \emph{IEEE Transactions on Industrial Informatics}, 2024.

\bibitem{wang2023transfer}
K.~Wang, H.~Wang, Z.~Yang, J.~Feng, Y.~Li, J.~Yang, and Z.~Chen, ``A transfer learning method for electric vehicles charging strategy based on deep reinforcement learning,'' \emph{Applied Energy}, vol. 343, p. 121186, 2023.

\bibitem{mahdian2012online}
M.~Mahdian, H.~Nazerzadeh, and A.~Saberi, ``Online optimization with uncertain information,'' \emph{ACM Transactions on Algorithms (TALG)}, vol.~8, no.~1, pp. 1--29, 2012.

\bibitem{banerjee2020improving}
S.~Banerjee, ``Improving online rent-or-buy algorithms with sequential decision making and ml predictions,'' \emph{Advances in Neural Information Processing Systems}, vol.~33, pp. 21\,072--21\,080, 2020.

\bibitem{rohatgi2020near}
D.~Rohatgi, ``Near-optimal bounds for online caching with machine learned advice,'' in \emph{Proceedings of the Fourteenth Annual ACM-SIAM Symposium on Discrete Algorithms}.\hskip 1em plus 0.5em minus 0.4em\relax SIAM, 2020, pp. 1834--1845.

\bibitem{lykouris2021competitive}
T.~Lykouris and S.~Vassilvitskii, ``Competitive caching with machine learned advice,'' \emph{Journal of the ACM (JACM)}, vol.~68, no.~4, pp. 1--25, 2021.

\bibitem{im2022parsimonious}
S.~Im, R.~Kumar, A.~Petety, and M.~Purohit, ``Parsimonious learning-augmented caching,'' in \emph{International Conference on Machine Learning}.\hskip 1em plus 0.5em minus 0.4em\relax PMLR, 2022, pp. 9588--9601.

\bibitem{antoniadis2020online}
A.~Antoniadis, C.~Coester, M.~Elias, A.~Polak, and B.~Simon, ``Online metric algorithms with untrusted predictions,'' in \emph{International Conference on Machine Learning}.\hskip 1em plus 0.5em minus 0.4em\relax PMLR, 2020, pp. 345--355.

\bibitem{li2022robustness}
T.~Li, R.~Yang, G.~Qu, G.~Shi, C.~Yu, A.~Wierman, and S.~Low, ``Robustness and consistency in linear quadratic control with untrusted predictions,'' \emph{ACM SIGMETRICS Performance Evaluation Review}, vol.~50, no.~1, pp. 107--108, 2022.

\bibitem{anand2022online}
K.~Anand, R.~Ge, A.~Kumar, and D.~Panigrahi, ``Online algorithms with multiple predictions,'' in \emph{International Conference on Machine Learning}.\hskip 1em plus 0.5em minus 0.4em\relax PMLR, 2022, pp. 582--598.

\bibitem{christianson2022chasing}
N.~Christianson, T.~Handina, and A.~Wierman, ``Chasing convex bodies and functions with black-box advice,'' in \emph{Conference on Learning Theory}.\hskip 1em plus 0.5em minus 0.4em\relax PMLR, 2022, pp. 867--908.

\bibitem{li2023beyond}
T.~Li, Y.~Lin, S.~Ren, and A.~Wierman, ``Beyond black-box advice: Learning-augmented algorithms for mdps with q-value predictions,'' \emph{Advances in Neural Information Processing Systems}, vol.~36, 2024.

\bibitem{lindermayr2023speed}
A.~Lindermayr, N.~Megow, and M.~Rapp, ``Speed-oblivious online scheduling: knowing (precise) speeds is not necessary,'' in \emph{International Conference on Machine Learning}.\hskip 1em plus 0.5em minus 0.4em\relax PMLR, 2023, pp. 21\,312--21\,334.

\bibitem{balkanski2024energy}
E.~Balkanski, N.~Perivier, C.~Stein, and H.-T. Wei, ``Energy-efficient scheduling with predictions,'' \emph{Advances in Neural Information Processing Systems}, vol.~36, 2024.

\bibitem{rosolia2017learning}
U.~Rosolia and F.~Borrelli, ``Learning model predictive control for iterative tasks. a data-driven control framework,'' \emph{IEEE Transactions on Automatic Control}, vol.~63, no.~7, pp. 1883--1896, 2017.

\bibitem{karnchanachari2020practical}
N.~Karnchanachari, M.~I. Valls, D.~Hoeller, and M.~Hutter, ``Practical reinforcement learning for mpc: Learning from sparse objectives in under an hour on a real robot,'' in \emph{Learning for Dynamics and Control}.\hskip 1em plus 0.5em minus 0.4em\relax PMLR, 2020, pp. 211--224.

\bibitem{brunke2022safe}
L.~Brunke, M.~Greeff, A.~W. Hall, Z.~Yuan, S.~Zhou, J.~Panerati, and A.~P. Schoellig, ``Safe learning in robotics: From learning-based control to safe reinforcement learning,'' \emph{Annual Review of Control, Robotics, and Autonomous Systems}, vol.~5, pp. 411--444, 2022.

\bibitem{koller2018learning}
T.~Koller, F.~Berkenkamp, M.~Turchetta, and A.~Krause, ``Learning-based model predictive control for safe exploration,'' in \emph{2018 IEEE conference on decision and control (CDC)}.\hskip 1em plus 0.5em minus 0.4em\relax IEEE, 2018, pp. 6059--6066.

\bibitem{hansen2022temporal}
N.~Hansen, X.~Wang, and H.~Su, ``Temporal difference learning for model predictive control,'' in \emph{International Conference on Machine Learning, PMLR}, 2022.

\bibitem{cheng2019control}
R.~Cheng, A.~Verma, G.~Orosz, S.~Chaudhuri, Y.~Yue, and J.~Burdick, ``Control regularization for reduced variance reinforcement learning,'' in \emph{International Conference on Machine Learning}.\hskip 1em plus 0.5em minus 0.4em\relax PMLR, 2019, pp. 1141--1150.

\bibitem{lin2021perturbation}
Y.~Lin, Y.~Hu, G.~Shi, H.~Sun, G.~Qu, and A.~Wierman, ``Perturbation-based regret analysis of predictive control in linear time varying systems,'' \emph{Advances in Neural Information Processing Systems}, vol.~34, pp. 5174--5185, 2021.

\bibitem{zhang2021regret}
R.~Zhang, Y.~Li, and N.~Li, ``On the regret analysis of online {LQR} control with predictions,'' in \emph{2021 American Control Conference (ACC)}.\hskip 1em plus 0.5em minus 0.4em\relax IEEE, 2021, pp. 697--703.

\bibitem{lin2022bounded}
Y.~Lin, Y.~Hu, G.~Qu, T.~Li, and A.~Wierman, ``Bounded-regret mpc via perturbation analysis: Prediction error, constraints, and nonlinearity,'' \emph{Advances in Neural Information Processing Systems}, vol.~35, pp. 36\,174--36\,187, 2022.

\bibitem{tang2020off}
Z.~Tang, Y.~Feng, N.~Zhang, J.~Peng, and Q.~Liu, ``Off-policy interval estimation with lipschitz value iteration,'' \emph{Advances in Neural Information Processing Systems}, vol.~33, pp. 7887--7897, 2020.

\bibitem{ghosh2022provably}
A.~Ghosh, X.~Zhou, and N.~Shroff, ``Provably efficient model-free constrained rl with linear function approximation,'' \emph{Advances in Neural Information Processing Systems}, vol.~35, pp. 13\,303--13\,315, 2022.

\bibitem{golowich2022can}
N.~Golowich and A.~Moitra, ``Can q-learning be improved with advice?'' in \emph{Conference on Learning Theory}.\hskip 1em plus 0.5em minus 0.4em\relax PMLR, 2022, pp. 4548--4619.

\bibitem{brockman2016openai}
G.~Brockman, V.~Cheung, L.~Pettersson, J.~Schneider, J.~Schulman, J.~Tang, and W.~Zaremba, ``Openai gym,'' \emph{arXiv preprint arXiv:1606.01540}, 2016.

\bibitem{silver2014deterministic}
D.~Silver, G.~Lever, N.~Heess, T.~Degris, D.~Wierstra, and M.~Riedmiller, ``Deterministic policy gradient algorithms,'' in \emph{International conference on machine learning}.\hskip 1em plus 0.5em minus 0.4em\relax Pmlr, 2014, pp. 387--395.

\end{thebibliography}

\end{document}